\documentclass[twocolumn]{aastex63}
\usepackage{mathptmx}
\usepackage{graphicx} 
\usepackage{threeparttable} 
\usepackage[caption=false]{subfig}
\usepackage{graphicx}
\usepackage{multirow}
\usepackage{graphicx}
\usepackage{booktabs}
\usepackage{amsmath, amsthm, amssymb, amsfonts}
\usepackage[T1]{fontenc}
\usepackage{bm}
\graphicspath{{./}{figures/}}

\def\degr{^\circ}

\begin{document}

\title{X-ray Polarization of the High-Synchrotron-Peaked BL Lac H 1426+428}
\correspondingauthor{Jin Zhang}
\email{j.zhang@bit.edu.cn}

\author{Xin-Ke Hu}
\affiliation{School of Physics, Beijing Institute of Technology, Beijing 100081, People's Republic of China; j.zhang@bit.edu.cn}

\author[0009-0003-9471-4724]{Jia-Xuan Li}
\affiliation{School of Physics, Beijing Institute of Technology, Beijing 100081, People's Republic of China; j.zhang@bit.edu.cn}

\author[0009-0000-6577-1488]{Yu-Wei Yu}
\affiliation{School of Physics, Beijing Institute of Technology, Beijing 100081, People's Republic of China; j.zhang@bit.edu.cn}

\author[0000-0003-2547-1469]{Ji-Shun Lian}
\affiliation{School of Physics, Beijing Institute of Technology, Beijing 100081, People's Republic of China; j.zhang@bit.edu.cn}

\author[0000-0002-9370-4079]{Wei Deng}
\affiliation{Guangxi Key Laboratory for Relativistic Astrophysics, School of Physical Science and Technology, Guangxi University, Nanning 530004, People's Republic of China}

\author[0000-0001-6863-5369]{Hai-Ming Zhang}
\affiliation{Guangxi Key Laboratory for Relativistic Astrophysics, School of Physical Science and Technology, Guangxi University, Nanning 530004, People's Republic of China}

\author[0000-0003-3554-2996]{Jin Zhang\dag}
\affiliation{School of Physics, Beijing Institute of Technology, Beijing 100081, People's Republic of China; j.zhang@bit.edu.cn}

\begin{abstract}

We report the X-ray polarization properties of the high-synchrotron-peaked BL Lac H 1426+428, based on two-epoch observational data from the Imaging X-ray Polarimetry Explorer (IXPE). For the first observation, only an upper limit of polarization degree ($\Pi_{\rm X}$), $\Pi_{\rm X}<19.5\%$, at the 99\% confidence level (C.L.) is determined. In contrast, for the second observation, we derive $\Pi_{\rm X}=20.6\%\pm2.9\%$ with a polarization angle ($\psi_{\rm X}$) of $\psi_{\rm X}=116.1\degr\pm4.1\degr$ at a C.L. of 7.1 $\sigma$. The time-resolved and energy-resolved polarization analysis reveals no significant variation in $\psi_{\rm X}$ and no detectable polarization within narrower energy bins for the first observation, while the polarization during the second observation is dominated by low-energy photons. Furthermore, the X-rays during the second observation are found to be in a higher flux state with a harder spectrum compared to that observed during the first observation, consistent with a {\it harder-when-brighter} behavior. We propose that the enhanced X-ray emission observed during the second observation is produced by shock-accelerated electrons within an ordered magnetic field region via synchrotron radiation. Nonetheless, no significant detection of polarization during the first IXPE observation may be due to the limited number of detected photons.

\end{abstract}


\section{Introduction}

Blazars are an exceptionally active subclass of active galactic nuclei (AGNs), characterized by their relativistic plasma jets that are nearly aligned with the observer's line of sight \citep{1995PASP..107..803U}. Their emission spans almost the entire electromagnetic spectrum, predominantly featuring non-thermal radiation from the jet. The broadband spectral energy distributions (SEDs) of blazars generally exhibit a bimodal shape. The first peak is attributed to synchrotron radiation, while the origin of the second peak remains a subject of debate. This second peak may arise from inverse-Compton (IC) scattering processes, including synchrotron self-Compton process \citep[SSC;][]{1992ApJ...397L...5M,1996A&AS..120C.503G,1996MNRAS.280...67G,2009ApJ...704...38S,2012ApJ...752..157Z,2014MNRAS.439.2933Y} and external Compton process \citep[EC;][]{1993ApJ...416..458D,1994ApJ...421..153S,2000ApJ...545..107B,2009MNRAS.397..985G,2014ApJ...788..104Z,2015ApJ...807...51Z}, or it could be explained by proton synchrotron radiation \citep{2000NewA....5..377A,2013ApJ...768...54B} or photo-pion production \citep{1993A&A...269...67M,2022ApJ...925L..19C}. Based on the peak frequency ($\nu_{\rm s,~p}$) of the synchrotron bump, blazars are classified into three subclasses \citep{2010ApJ...710.1271A,2016ApJS..226...20F}: low-synchrotron-peaked blazars (LSPs, $\nu_{\rm s,~p}<10^{14}~{\rm Hz}$), intermediate-synchrotron-peaked blazars (ISPs, $10^{14}~{\rm Hz}<\nu_{\rm s,~p}<10^{15}~{\rm Hz}$) and high-synchrotron-peaked blazars (HSPs, $\nu_{\rm s,~p}>10^{15}~{\rm Hz}$).

To investigate the acceleration mechanisms of relativistic particles in blazar jets represents one of the most compelling challenges in contemporary astrophysics \citep{2021Galax...9...37T}. The primary acceleration processes encompass shock acceleration \citep{1987PhR...154....1B}, magnetic reconnection \citep{2014ApJ...783L..21S}, and stochastic acceleration \citep{2018PhRvL.121y5101C}. Polarization measurements serve as a critical tool for distinguishing between various radiation mechanisms \citep{2013ApJ...768...54B} and for probing the underlying acceleration processes \citep{2018MNRAS.480.2872T,2021Galax...9...37T}. While previous polarization observations have been confined to optical and radio wavelengths, X-ray polarimetry is particularly vital because the emission region of X-ray photons is more proximate to the particle acceleration site \citep{2022ApJ...938L...7D}. The Imaging X-ray Polarimetry Explorer \citep[IXPE;][]{2022JATIS...8b6002W}, launched on 2021 December 9, represents the first telescope capable of conducting the polarization measurements of blazars at X-ray energies. IXPE has observed more than 10 blazars, and all the HSPs among them, such as Mrk 501 \citep{2022Natur.611..677L,2024ApJ...970L..22H}, Mrk 421 \citep{2022ApJ...938L...7D,2024A&A...681A..12K}, PG 1553+113 \citep{2023ApJ...953L..28M}, 1ES 0229+200 \citep{2023ApJ...959...61E}, 1ES 1959+650 \citep{2024ApJ...963L..41H,2024ApJ...963....5E} and PKS 2155-304 \citep{2024ApJ...963L..41H,2024A&A...689A.119K}, exhibit significant polarization in the 2--8 keV band, supporting the presence of ordered magnetic fields in the X-ray emission regions of these sources. Moreover, the chromatic behaviors of polarization degree ($\Pi$) and the alignment of the multiwavelength polarization angles ($\psi$) with the jet direction found in some HSPs suggest that shock acceleration mechanism may play a crucial role in particle acceleration within these jets \citep{2022Natur.611..677L,2022ApJ...938L...7D,2023ApJ...953L..28M,2023ApJ...959...61E,2024A&A...681A..12K,2024ApJ...970L..22H,2024A&A...689A.119K}.

H 1426+428 is a TeV-emitting HSP \citep{2002ApJ...571..753H,2002A&A...384L..23A,2002A&A...391L..25D}, located at $z=0.129$ \citep{2010AJ....139..390P}. Additionally, it belongs to the rare class of extreme-high-synchrotron-peaked blazars \citep[EHSPs, $\nu_{\rm s,~p}>10^{17}~{\rm Hz}$;][]{2001A&A...371..512C,2008ASPC..386..302W}. It has been detected by the Fermi-LAT during the first 5.5 months of observations, exhibiting weak $\gamma$-ray emission below 20 GeV with a photon spectral index of $\Gamma\sim1.5$ \citep{2009ApJ...707.1310A}. The observations with Beppo-SAX reveal that the synchrotron radiation from H 1426+428 can extend beyond 100 keV \citep{2001A&A...371..512C}. The X-ray emission from this source exhibits significant variability on both short (days) and long (weeks) timescales \citep{2009AstL...35..579F}. \citet{2023MNRAS.520.4118C} report that a potential quasi-periodic oscillation with a period of $P=48.67\pm13.90$ minutes was observed in the $I$-band observations of H 1426+428 on 2010 April 13. Additionally, a Very Long Baseline Array (VLBA) observation at 8 GHz on 2006 October 22 reveals a 3 mJy jet component approximately 2 mas northwest of the $\sim$ 20 mJy core \citep{2010ApJ...723.1150P}. In this study, we conduct the first X-ray polarimetric analysis of H 1426+428 using observational data from IXPE. Additionally, we perform a comprehensive multiwavelength analysis using data obtained from Fermi-LAT \citep{2009ApJ...697.1071A}, NuSTAR \citep{2013ApJ...770..103H}, and Swift-XRT \citep{2004ApJ...611.1005G,2005SSRv..120..165B}.

\section{IXPE Observations and Data Reduction}\label{sec_ixpe}

H 1426+428 was observed by IXPE in two different periods: 2024 May 27--29 with a total exposure of $\sim$ 106 ks (OBSID: 03007201) and 2024 July 5--7 with a total exposure of $\sim$ 104 ks (OBSID: 03007301). Both the first and second IXPE observational data are analyzed in this manuscript.

Coordinate correction is applied to the publicly available Level-2 event files to correct for detector pointing misalignment. Given that H 1426+428 is a point-like source with an angular resolution of $\sim30^{\prime\prime}$ as observed by IXPE, and considering that the X-ray emission can extend up to $60^{\prime\prime}$, the source region is defined as a circle centering on the coordinates of the light centroid as measured on the Hubble Space Telescope (HST) image \citep[i.e., R.A. = $217\degr.136$, Decl. = $42\degr.672$;][]{2000ApJ...532..740S} with a radius of $60^{\prime\prime}$. The background region is defined as an annulus with inner and outer radii of $120^{\prime\prime}$ and $270^{\prime\prime}$, which is concentric with the source region. This strategy for selecting the background region follows the recommendation provided in \citet{2023AJ....165..143D}. Source and background photons are extracted from their respective regions using the \texttt{xpselect} task within the software \texttt{ixpeobssim} \citep[v.31.0.3;][]{2022SoftX..1901194B}. Two different methods are employed to estimate the polarization properties of H 1426+428: a model-independent method implemented in the \texttt{PCUBE} algorithm within the \texttt{xpbin} task of the \texttt{ixpeobssim} \citep{2015APh....68...45K} and a spectropolarimetric analysis using \texttt{Xspec} \citep[v.12.14.1;][]{1999ascl.soft10005A} as described in \citet{2017ApJ...838...72S}. For the \texttt{PCUBE} analysis, the influence of the sky background is excluded by applying the background subtraction procedure outlined in \citet{2022SoftX..1901194B}.

We analyze the data of the first and second IXPE observations for H 1426+428 with three combined detector units (DUs) using the \texttt{PCUBE} algorithm to derive polarization properties, e.g., the normalized Stokes parameters $q$ and $u$ ($q=Q/I$ and $u=U/I$), the minimum detectable polarization at the 99\% confidence level (C.L.; MDP$_{99}$), the X-ray polarization degree ($\Pi_{\rm X}$), the X-ray polarization angle ($\psi_{\rm X}$) and their associated 1$\sigma$ uncertainties. The \texttt{PCUBE} analysis is performed using the unweighted analysis method \citep[i.e., weights=False and irfname="ixpe:obssim:20240701:v013";][]{2022AJ....163..170D}.

Spectropolarimetric analysis is performed using the \texttt{Xspec} software within the \texttt{HEASoft} package (v.6.34). The Stokes parameter $I$, $Q$, and $U$ spectra are generated using the \texttt{PHA1}, \texttt{PHA1Q}, and \texttt{PHA1U} algorithms, respectively, within the \texttt{xpbin} task. A weighted analysis method \citep[i.e., weights=True and irfname="ixpe:obssim:20240701$\_$alpha075:v013";][]{2022AJ....163..170D} with the \texttt{alpha075} response matrix files (RMFs) is employed for the $I$, $Q$, and $U$ spectra to enhance measurement significance. The $I$ spectra are grouped to ensure a minimum of 20 counts per energy bin, as required for $\chi^{2}$ statistics during spectral fits. Meanwhile, both the $Q$ and $U$ spectra are binned with a constant energy width of 0.2 keV. We fit these spectra following a two-step procedure \citep{2024ApJ...962...92X,2024ApJ...963....5E,2024ApJ...970L..22H}. Firstly, the $I$ spectra are jointly fitted with the quasi-simultaneous Swift-XRT spectra (see Section \ref{sec_xrt} in the Appendix for details) using an absorbed power-law (PL) model, i.e., \texttt{CONSTANT}$\times$\texttt{TBABS}$\times$\texttt{POWERLAW} within \texttt{Xspec}. The PL function is
\begin{equation}
\frac{dN}{dE}=N_{0}\times\left(\frac{E}{E_0}\right)^{-\Gamma_{\rm X}},
\end{equation}
where $N_0$ is the PL normalization, $E_{0}=1~{\rm keV}$ is the scale parameter of photon energy, and $\Gamma_{\rm X}$ is the photon spectral index \citep{2004A&A...413..489M}. The \texttt{CONSTANT} and \texttt{TBABS} models take into account the uncertainties in the absolute effective areas of different detectors and the Galactic photoelectric absorption, respectively. The joint spectral fits enable more precise constraints on the column density ($N_{\rm H}$) and the spectral shape \citep{2024ApJ...963....5E}. Secondly, we simultaneously fit the $I$, $Q$ and $U$ spectra from all three DUs using an absorbed PL model with constant polarization, i.e., \texttt{CONSTANT}$\times$\texttt{TBABS}$\times$\texttt{POLCONST}$\times$\texttt{POWERLAW} within \texttt{Xspec}. The polarization model \texttt{POLCONST} assumes a constant polarization within the operating energy range and includes only two free parameters, namely $\Pi_{\rm X}$ and $\psi_{\rm X}$. During the spectropolarimetric fits, all other spectral parameters are fixed at their best-fit values obtained from the first-step analysis, while $\Pi_{\rm X}$ and $\psi_{\rm X}$ are allowed to vary freely.

\section{Results of the X-ray Polarization Analysis}\label{sec_results}
\subsection{Time- and Energy-Averaged Analysis}

As described in Section \ref{sec_ixpe}, we utilize two different methods for the analysis of IXPE observational data of H 1426+428 as a cross-check: a model-independent approach using the \texttt{PCUBE} algorithm and a spectropolarimetric analysis method using \texttt{Xspec}.

Through the \texttt{PCUBE} analysis, no significant polarization is detected in the 2--8 keV band during the first IXPE observation. The normalized Stokes parameters are measured as $q=0.086\pm0.057$ and $u=0.056\pm0.059$, with MDP$_{99}$ of $\sim17.9\%$. However, during the second IXPE observation, a polarization is detected at a C.L. of 4.8$\sigma$, yielding $\Pi_{\rm X}=19.3\%\pm4.5\%$ and $\psi_{\rm X}=118.3\degr\pm6.7\degr$ with MDP$_{99}$ of $\sim13.9\%$. The normalized Stokes parameters for the two IXPE observations are presented in Figure \ref{fig_qu}. The $\Pi_{\rm X}$ value of H 1426+428 in the 2--8 keV band during the second IXPE observation is similar to that of some other HSPs observed by IXPE, such as 1ES 0229+200 \citep{2023ApJ...959...61E} and PKS 2155--304 \citep{2024ApJ...963L..41H,2024A&A...689A.119K}.

In the spectropolarimetric analysis, the joint spectral fits yield $\Gamma_{\rm X}=2.29\pm0.04$ and $\Gamma_{\rm X}=2.00\pm0.03$ for the first and second IXPE observations, respectively. The unabsorbed fluxes of H 1426+428 in the 2--8 keV band ($F_{\rm 2-8~keV}$) for these two epochs are also estimated through joint spectral fits using the \texttt{cflux} task within \texttt{Xspec}, namely $F_{\rm 2-8~keV}=(1.85\pm0.03)\times10^{-11}~{\rm erg~cm^{-2}~s^{-1}}$ for the first IXPE observation and $F_{\rm 2-8~keV}=(3.21\pm0.04)\times10^{-11}~{\rm erg~cm^{-2}~s^{-1}}$ for the second IXPE observation, respectively. Through the spectropolarimetric fits, an upper limit of $\Pi_{\rm X}<19.5\%$ at 99\% C.L. is derived in the 2--8 keV band for the first IXPE observation, with $\psi_{\rm X}$ remaining unconstrained. For the second IXPE observation, an X-ray polarization of $\Pi_{\rm X}=20.6\%\pm2.9\%$ with $\psi_{\rm X}=116.1\degr\pm4.1\degr$ is detected at 7.1$\sigma$ C.L.. The best-fit parameters of the spectropolarimetric fits are summarized in Table \ref{tab_specpol}, and the spectropolarimetric spectra are presented in Figure \ref{fig_specpol}.

It is observed that the X-ray polarization parameters of H 1426+428 obtained from the two analysis methods are consistent within their respective uncertainties. However, minor discrepancies exist between the best-estimated values of $\Pi_{\rm X}$ and $\psi_{\rm X}$ derived from these two methods. This can be attributed to the fact that the \texttt{PCUBE} analysis provides model-independent polarization parameters, whereas spectropolarimetric fits estimate polarization parameters while accounting for the best-fit spectral shape modeling \citep{2024A&A...681A..12K}. Given that spectropolarimetric fits can enhance sensitivity through the event weight method (details provided in \citealp{2022AJ....163..170D}), the subsequent discussion on the X-ray polarization of H 1426+428 will be based on the results from the spectropolarimetric analysis.

\subsection{Time-Resolved Analysis}

To investigate the temporal variations in X-ray polarization, particularly to determine if the rotation of $\psi_{\rm X}$ could account for the week polarization observed in H 1426+428 during the first IXPE observation, we conduct a time-resolved polarization analysis on the IXPE observational data following the method described in \citet{2024A&A...681A..12K} and \citet{2024ApJ...970L..22H}. We assess the time variability of X-ray polarization by calculating the null hypothesis probability ($P_{\rm Null}$) using a $\chi^{2}$ test, which assumes that the X-ray polarization remains constant throughout each IXPE observation period. The polarization parameters are estimated using the \texttt{PCUBE} algorithm.

Firstly, the individual IXPE observation is divided into identical time intervals depending on the selected bin number, and then we calculate $q$ and $u$ for each time interval. Secondly, we compare $q$ and $u$ with the results from fitting each parameter as constant during the observation time. Finally, we calculate the $\chi^{2}$ and $P_{\rm Null}$ values for each case. Both the first and second IXPE observations are segmented into two to six time intervals. $P_{\rm Null}$ as a function of the time interval number for both IXPE observations is shown in Figure \ref{fig_pnull}. $P_{\rm Null}>1\%$ indicates that the X-ray polarization remains statistically constant, whereas $P_{\rm Null}<1\%$ suggests that temporal variability exists in the X-ray polarization \citep{2024A&A...681A..12K,2024ApJ...970L..22H}. As depicted in Figure \ref{fig_pnull}, the splitting of the $q$(t) and $u$(t) light curves exhibit satisfactory fits with the constant model across all time intervals. Therefore, no significant temporal variability in X-ray polarization is detected during both the first and second IXPE observations of H 1426+428.

\subsection{Energy-Resolved Analysis}

To investigate potential energy-dependent variations in the X-ray polarization of H 1426+428, we partition the full 2--8 keV IXPE band into smaller, equal-width energy bins and estimate the X-ray polarization parameters for each energy bin. Both the \texttt{PCUBE} analysis and spectropolarimetric analysis are employed. During spectropolarimetric fits, the spectral parameters are fixed at the best-fit values obtained from the joint IXPE and Swift-XRT spectral fits, where the Swift-XRT observations are quasi-simultaneous with these IXPE observations, while only $\Pi_{\rm X}$ and $\psi_{\rm X}$ are allowed to vary. This procedure is conducted for two energy intervals (i.e., 2--5 and 5--8 keV), three intervals (i.e., 2--4, 4--6, and 6--8 keV), and up to six intervals.

For the first IXPE observation, no significant X-ray polarization is detected in any energy bins, with the estimated $\Pi_{\rm X}$ remaining below the corresponding MDP$_{99}$. For the second IXPE observation, X-ray polarization is detected at a C.L. exceeding 99\% in the 2--5 keV, 2--4 keV, 2--3 keV, and 3--4 keV energy bands, as shown in Figure \ref{fig_ebin}. The derived highest polarization degree is $\Pi_{\rm X}=25.8\%\pm5.5\%$ with $\psi_{\rm X}=111.7\degr\pm6.2\degr$ at a C.L. of 4.7$\sigma$ in the 3--4 keV band. This pattern of detecting X-ray polarization in lower energy bands but not in higher ones is similar to observations made in BL Lacertae \citep{2023ApJ...948L..25P} and PKS 2155-304 \citep{2024ApJ...963L..41H}.

\section{Flux and Spectral Variations}\label{sec_var}
\subsection{X-rays}

Through the analysis of IXPE observational data, we find that the source exhibits different flux states and spectral indices during the two observations. To investigate the X-ray flux states of H 1426+428 during the IXPE observations, we construct the long-term X-ray light curve using the data obtained from the Swift-XRT monitoring program for Fermi-LAT sources of interest \footnote{\url{https://www.swift.psu.edu/monitoring/}} \citep{2013ApJS..207...28S}. These data have been processed to include background subtraction, point-spread function (PSF) corrections, and pile-up corrections, as illustrated in Figure \ref{fig_lc}. It is worth noting that the flux light curve presented in Figure \ref{fig_lc} has been converted from the original count rate light curve using the conversion factor provided by \citet{2013ApJS..207...28S}. The X-ray emission of H 1426+428 shows significant variability, with a C.L. significantly exceeding 5$\sigma$, as assessed via $\chi^{2}$ test. As indicated by the magenta and orange horizontal dashed lines in Figure \ref{fig_lc}, the X-ray flux of H 1426+428 is in a moderate state during the first IXPE observation and shows a slight increase during the second one.

Additionally, to further explore the spectral variations of X-rays for H 1426+428, a comprehensive analysis is conducted using data from two NuSTAR observations combined with quasi-simultaneous Swift-XRT observations, as well as Swift-XRT observations which were carried out contemporaneously with two observations performed by the Major Atmospheric Gamma-ray Imaging Cherenkov telescopes \citep[MAGIC;][]{2009arXiv0907.0959L,2020ApJS..247...16A}. The details of data analysis for Swift-XRT and NuSTAR observations are described in Sections \ref{sec_xrt} and \ref{sec_nustar}. Using the analysis results, we present $\Gamma_{\rm X}$ as a function of $F_{\rm 2-10~keV}$ in Figure \ref{fig_flux-gamma}, where $F_{\rm 2-10~keV}$ represents the flux in the 2--10 keV band. Notably, a {\it harder-when-brighter} correlation tendency between $F_{\rm 2-10~keV}$ and $\Gamma_{\rm X}$ is observed on H 1426+428. Considering the uncertainties associated with both $\Gamma_{\rm X}$ and $F_{\rm 2-10~keV}$, we use the bootstrap method \citep{1979AnSta...7....1E} to estimate the correlation coefficient ($r$) between $\Gamma_{\rm X}$ and $F_{\rm 2-10~keV}$. This analysis results in $r=-0.76\pm0.05$, which indicates a significant tendency toward {\it harder-when-brighter} behavior in X-rays for H 1426+428. A similar trend has also been reported in other HSPs, such as Mrk 421 \citep{2022ApJ...938L...7D}, PKS 2155–304 \citep{2024ApJ...963L..41H}, and Mrk 501 \citep{2024ApJ...970L..22H}. This phenomenon is usually explained as the injection of high-energy electrons into the emission region \citep{1998A&A...333..452K,2022ApJ...938L...7D,2024ApJ...965...58Z}.

\subsection{Other Bands}

We generate a long-term light curve in the $\gamma$-ray band using Fermi-LAT observational data (see Section \ref{sec_lat} for details), as shown in Figure \ref{fig_lc}. No significant variability is detected in the GeV energy range, with a C.L. lower than 1$\sigma$ estimated through a $\chi^{2}$ test. Apparently, the GeV $\gamma$-ray flux remains stable when the X-rays exhibit significant variability. However, the photon spectral index ($\Gamma_\gamma$) demonstrates a degree of variation, with a minimum value of $\Gamma_\gamma=1.00\pm0.20$ and a maximum value of $\Gamma_\gamma=2.74\pm0.35$.

Using the data obtained from \citeauthor{NED}, MAGIC observations \citep{2009arXiv0907.0959L,2020ApJS..247...16A}, and the data analysis results of this work, we construct the broadband SED (details refer to Section \ref{sec_sed}) for H 1426+428, as illustrated in Figure \ref{fig_sed}. The SED exhibits a distinct bimodal structure, consistent with observations of other TeV BL Lacs \citep[e.g.,][]{2012ApJ...752..157Z}. The MAGIC observation conducted in 2008 yielded only upper limits \citep{2009arXiv0907.0959L}, while the X-ray emission is characterized by a low flux state and a soft spectrum during that period. In contrast, the TeV $\gamma$-rays of H 1426+428 were successfully detected by MAGIC in 2012 \citep{2020ApJS..247...16A}, when the X-rays were in a relatively higher flux state with a hard spectrum. This suggests a potential correlation between TeV and X-ray emission. Furthermore, it is observed that the synchrotron peak shifts to higher energies when the source exhibits enhanced X-ray and TeV emission states.

\section{Discussion and Conclusions}\label{sec_disc&conc}

The data analysis results reveal that significant X-ray polarization was detected during the second IXPE observation, with measured $\Pi_{\rm X}=20.6\%\pm2.9\%$ and $\psi_{\rm X}=116.1\degr\pm4.1\degr$, at a C.L. of 7.1$\sigma$ in the 2--8 keV band. This finding strongly suggests that the X-rays of H 1426+428 are generated via the synchrotron radiation within an emission region characterized by ordered magnetic fields. Conversely, for the first IXPE observation, only an upper limit of $\Pi_{\rm X}$ could be established, i.e., $\Pi_{\rm X}<19.5\%$ at 99\% C.L. in the 2--8 keV band. As discussed in \citet{2023NatAs...7.1245D}, the rapidly smooth rotation of $\psi_{\rm X}$ may lead to the non-detection of time-averaged polarization for Mrk 421. However, a time-resolved analysis of the IXPE observational data for H 1426+428 revealed no temporal variability in X-ray polarization in either observation, confirming that no rotation of $\psi_{\rm X}$ occurred during the first IXPE observation. Consequently, the lack of significant X-ray polarization detection during the first IXPE observation is unlikely to be explained by this mechanism.

\begin{figure*}
    \centering
    \includegraphics[angle=0, scale=0.35]{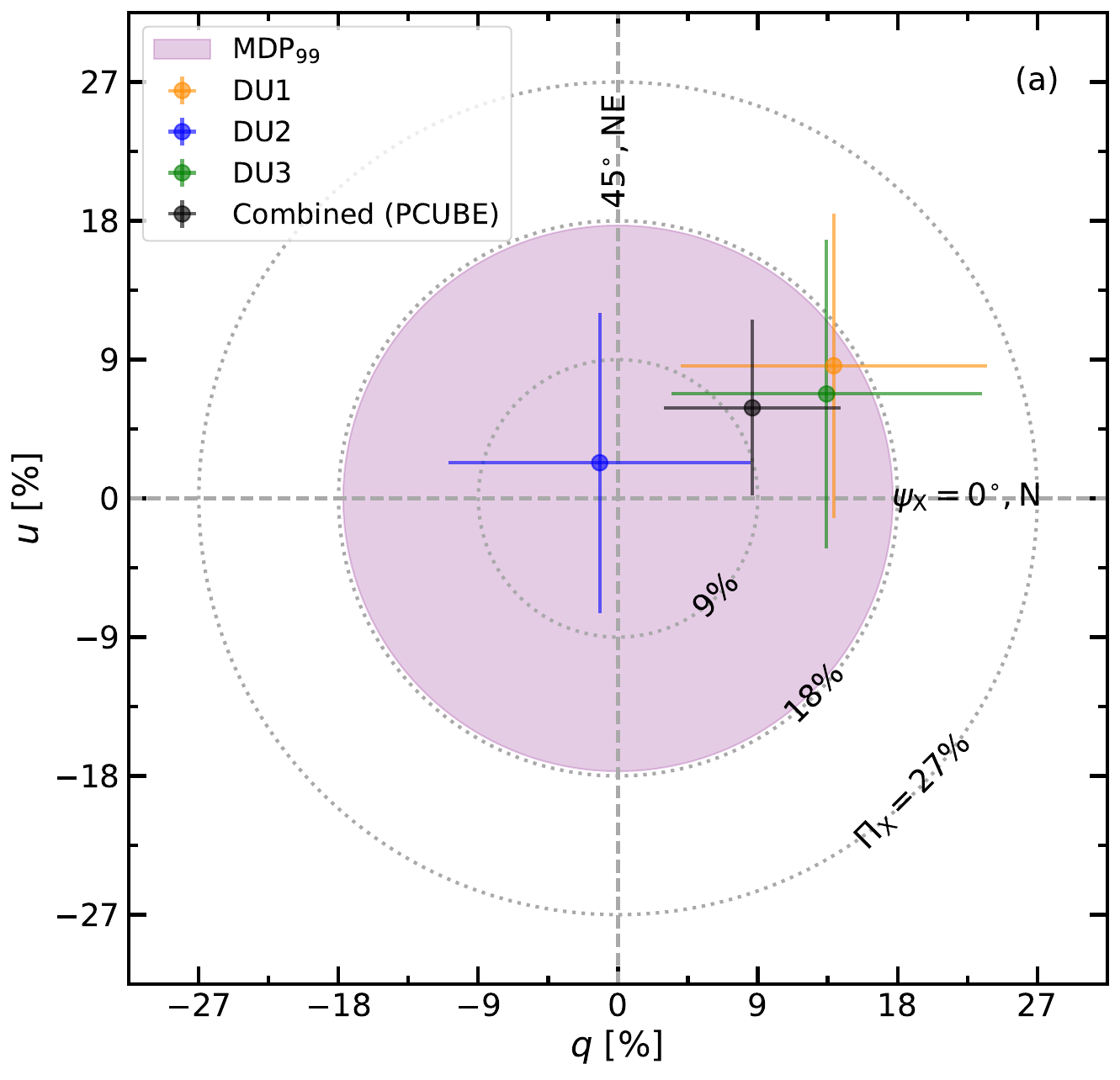}
    \includegraphics[angle=0, scale=0.35]{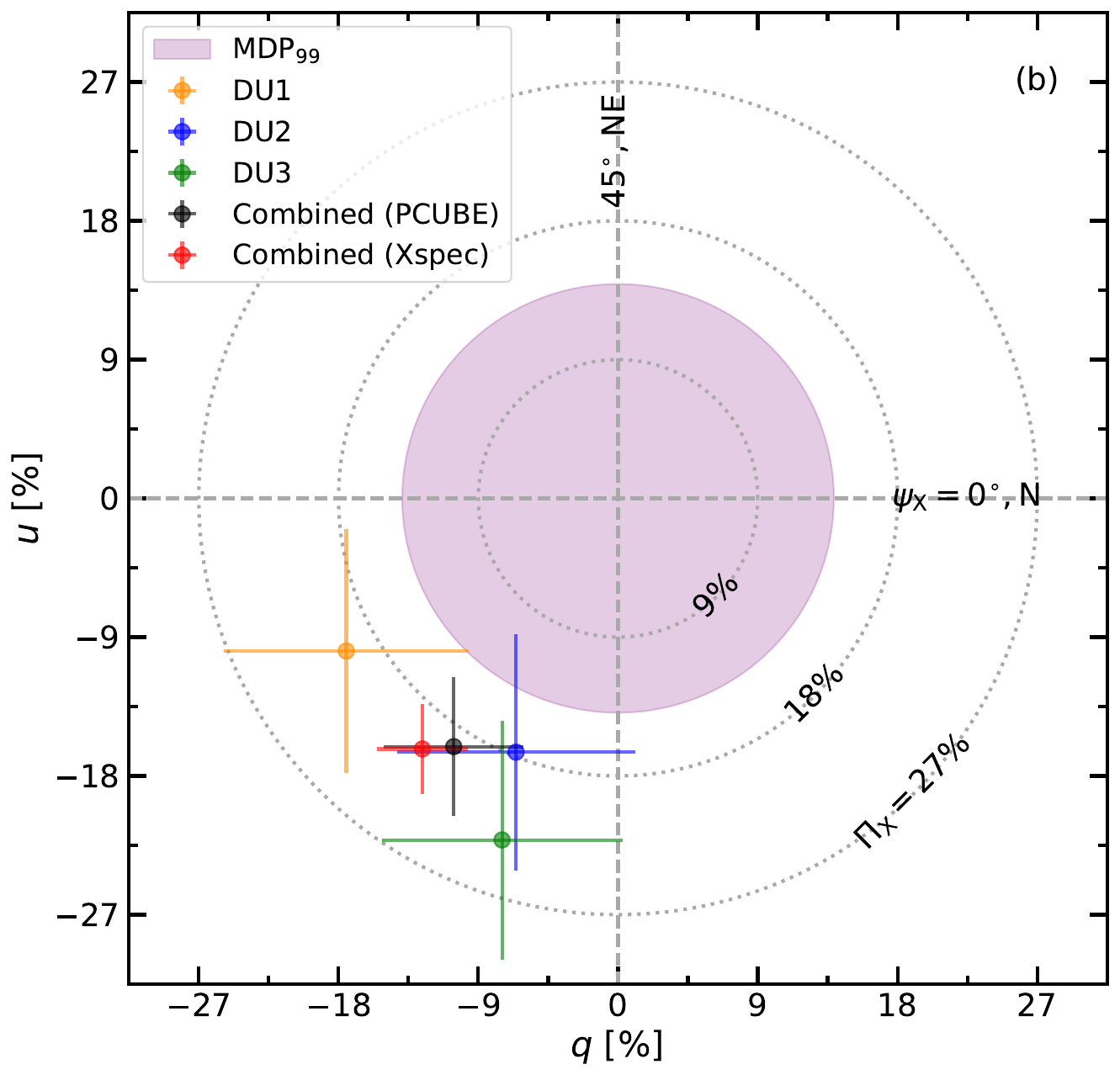}
    \caption{Normalized Stokes parameters $q$ and $u$ of the first ({\it panel} (a)) and second ({\it panel} (b)) IXPE observations for H 1426+428. The orange, blue, green, and black points represent the results of DU1, DU2, DU3, and the combination of the three DUs, respectively, which are measured with the \texttt{PCUBE} algorithm within \texttt{ixpeobssim}. The violet shaded areas represent the MDP$_{99}$ of the two IXPE observations for H 1426+428. The red point in {\it panel} (b) represents the result of the combination of the three DUs, estimated via spectropolarimetric fits within \texttt{Xspec}.}
    \label{fig_qu}
\end{figure*}

\begin{figure*}
    \centering
    \includegraphics[angle=0, scale=0.35]{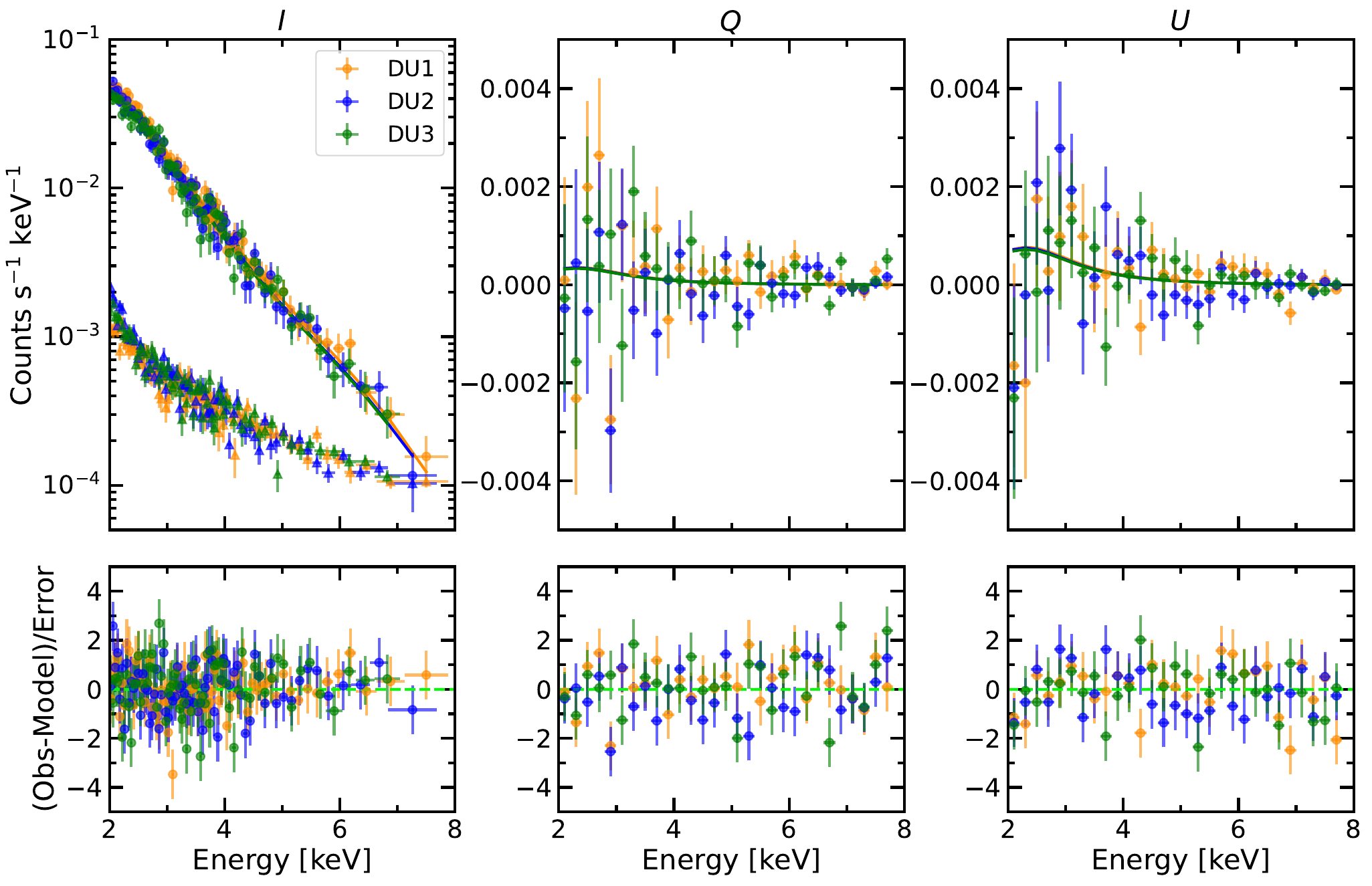}
    \includegraphics[angle=0, scale=0.35]{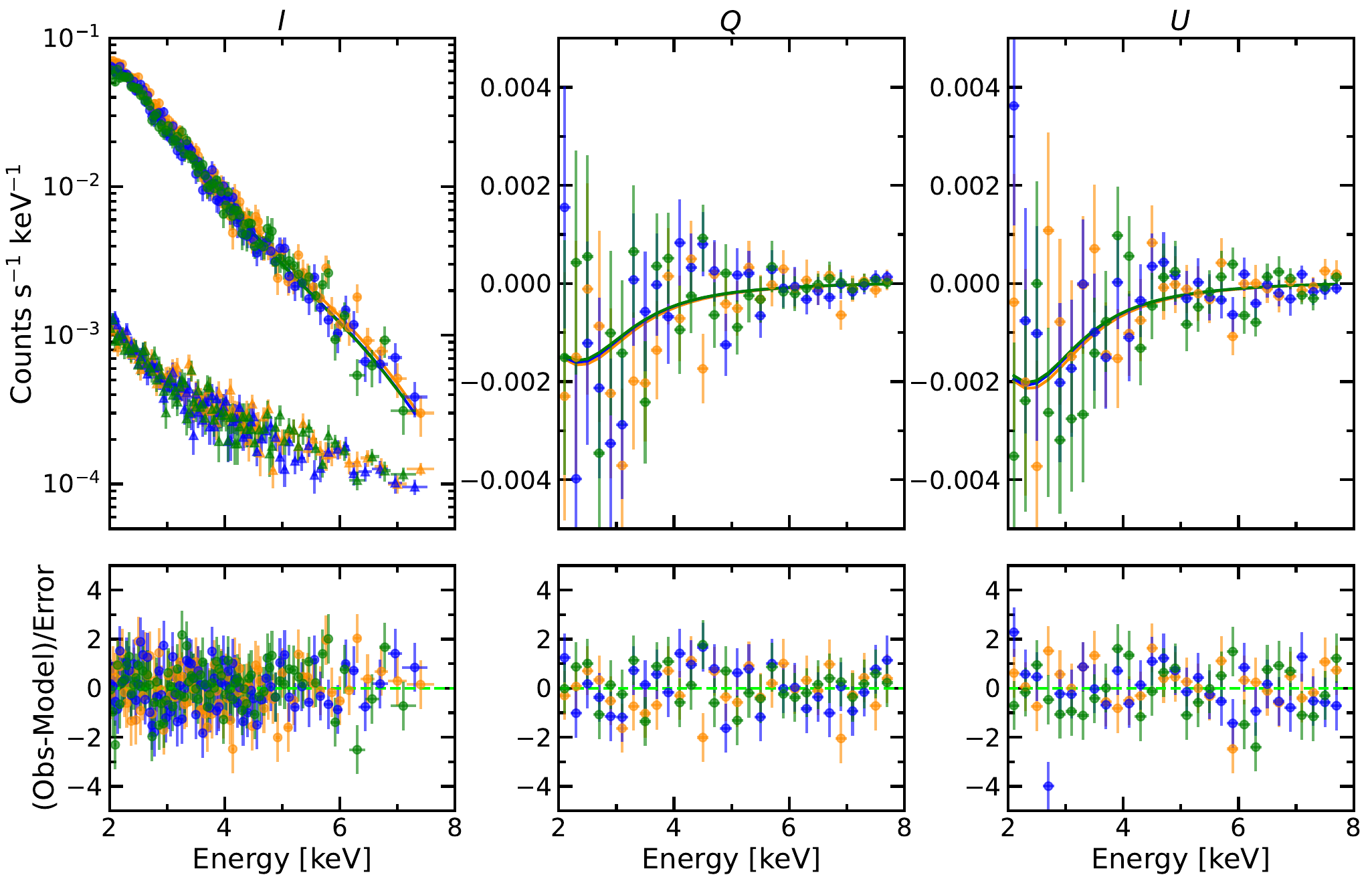}
    \caption{Spectropolarimetric fit results of the first ({\it top panels}) and second ({\it bottom panels}) IXPE observations for H 1426+428. Panels represent the fits to IPXE Stokes parameters $I$, $Q$ and $U$ with their associated residuals from left to right. In the $I$ spectra panels, the triangles indicate the background spectra for different DUs.}
    \label{fig_specpol}
\end{figure*}

\begin{figure*}
    \centering
    \includegraphics[angle=0, scale=0.45]{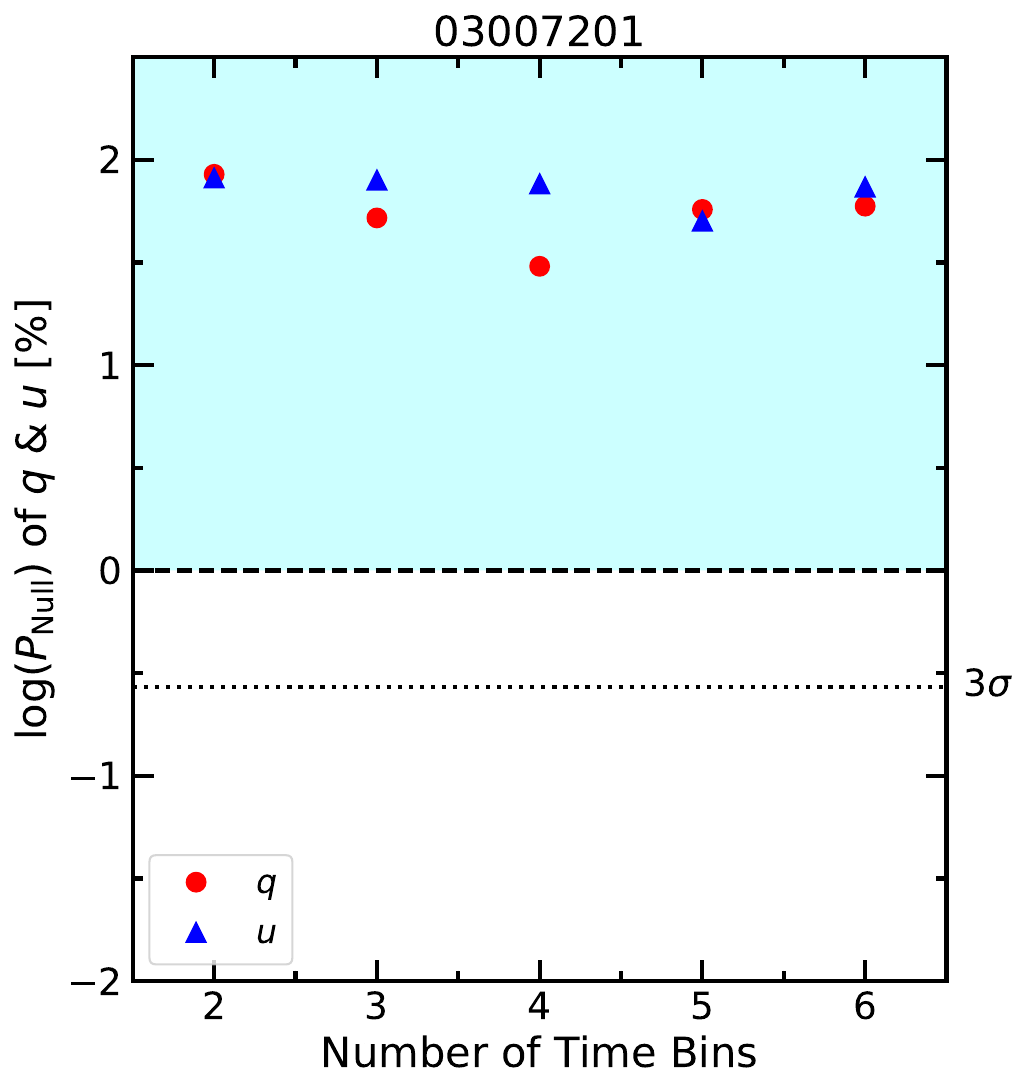}
    \includegraphics[angle=0, scale=0.45]{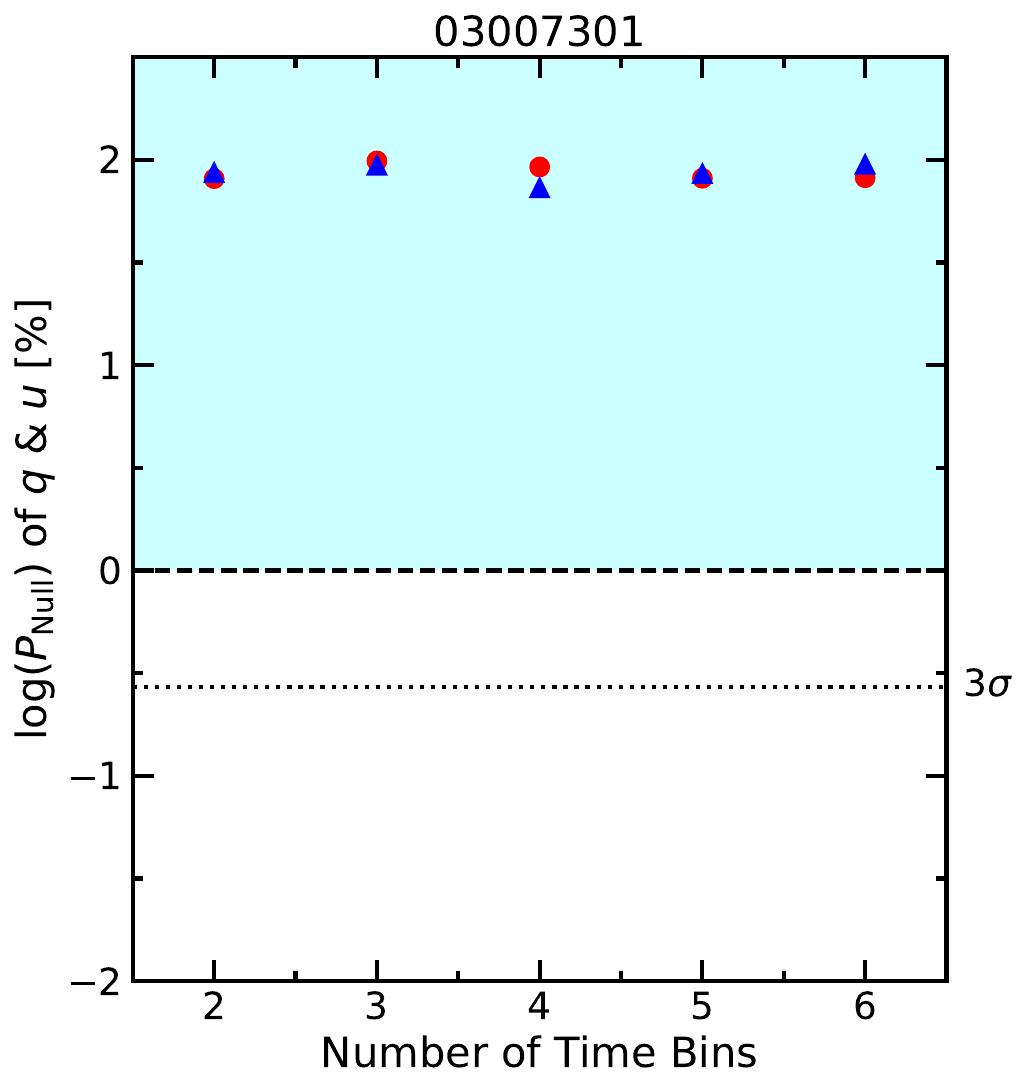}
    \caption{Null hypothesis probability ($P_{\rm Null}$) of the $\chi^{2}$ test, under the assumption that the X-ray polarization remains constant throughout each IXPE observation period, for the variability of the normalized Stokes parameters $q$ (red points) and $u$ (blue points), corresponding to the first ({\it left panel}) and second ({\it right panel}) IXPE observations of H 1426+428. The cyan shaded areas represent the regions with $P_{\rm Null}>1\%$ and the dotted lines mark the $3\sigma$ (99.73\%) significance level.}
    \label{fig_pnull}
\end{figure*}

\begin{figure*}
    \centering
    \includegraphics[angle=0, scale=0.25]{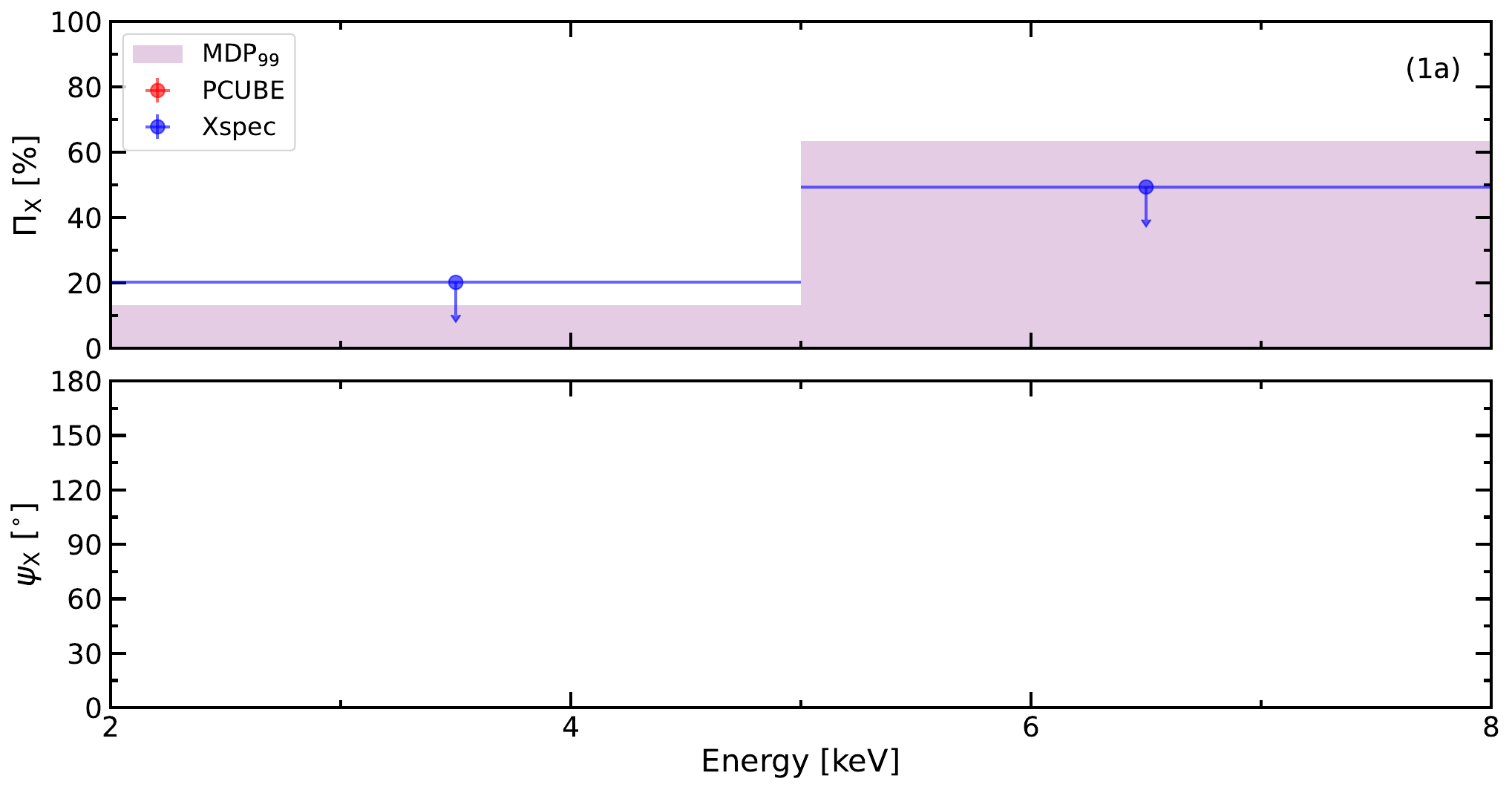}
    \includegraphics[angle=0, scale=0.25]{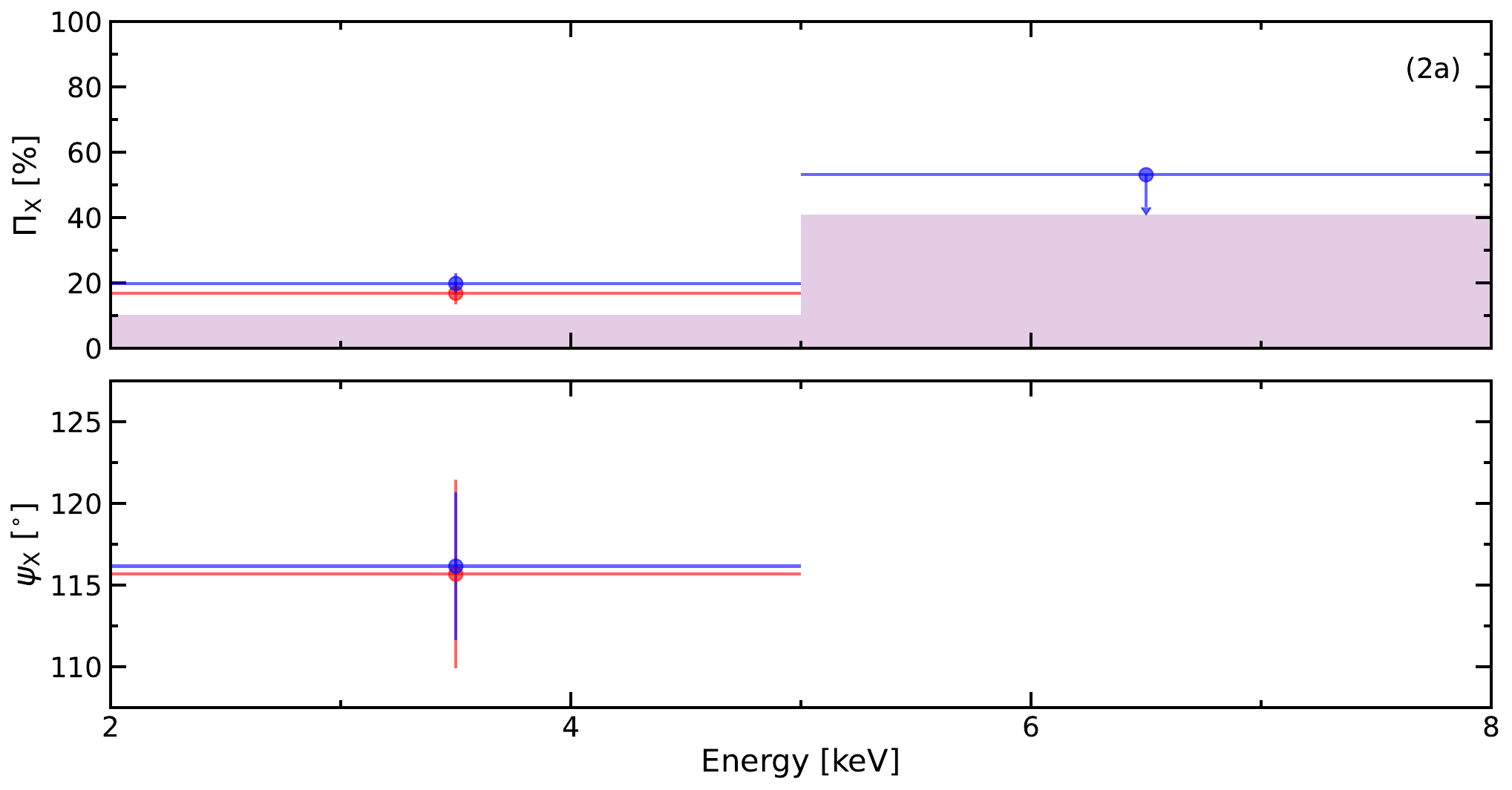}
    \includegraphics[angle=0, scale=0.25]{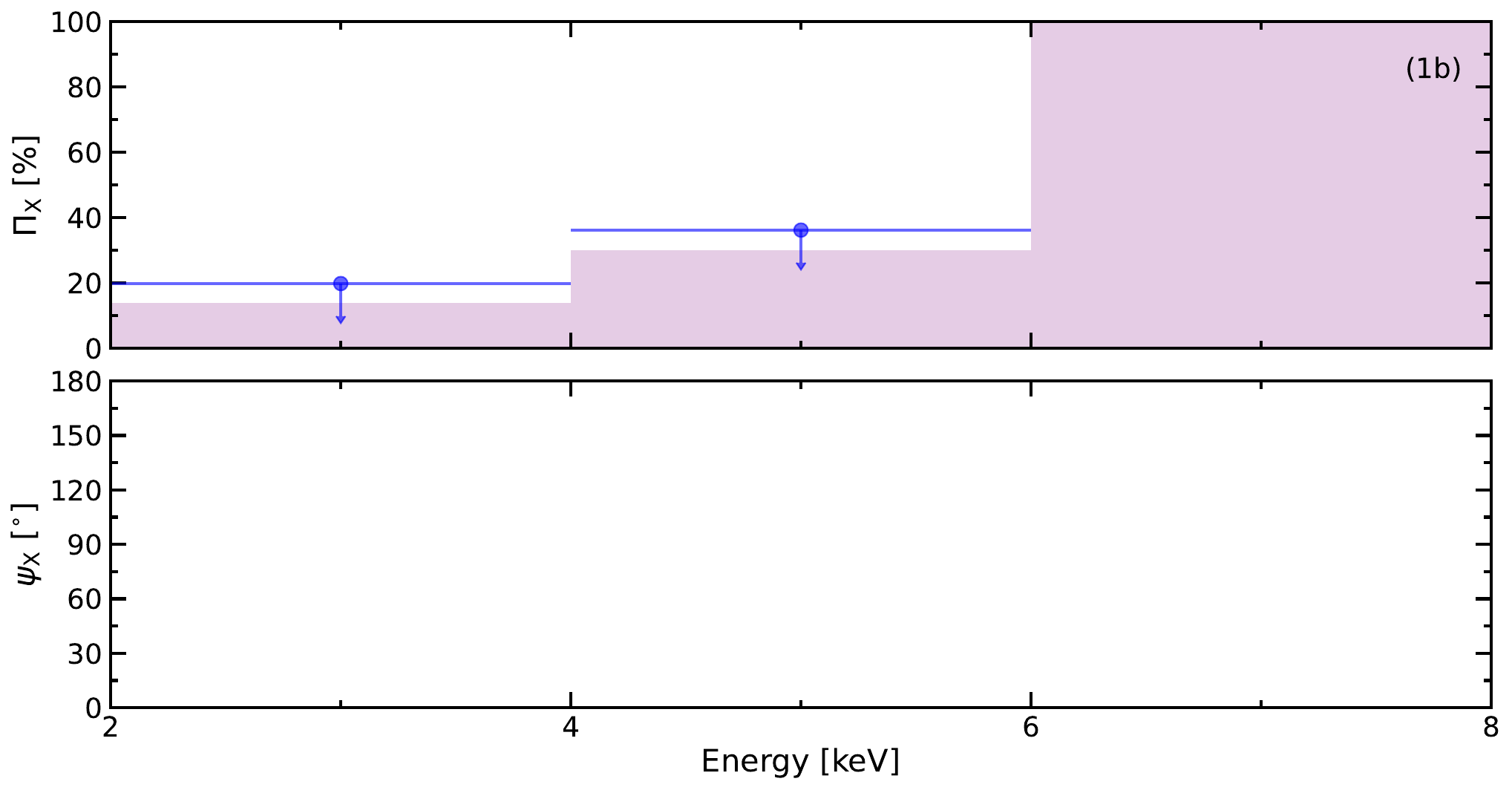}
    \includegraphics[angle=0, scale=0.25]{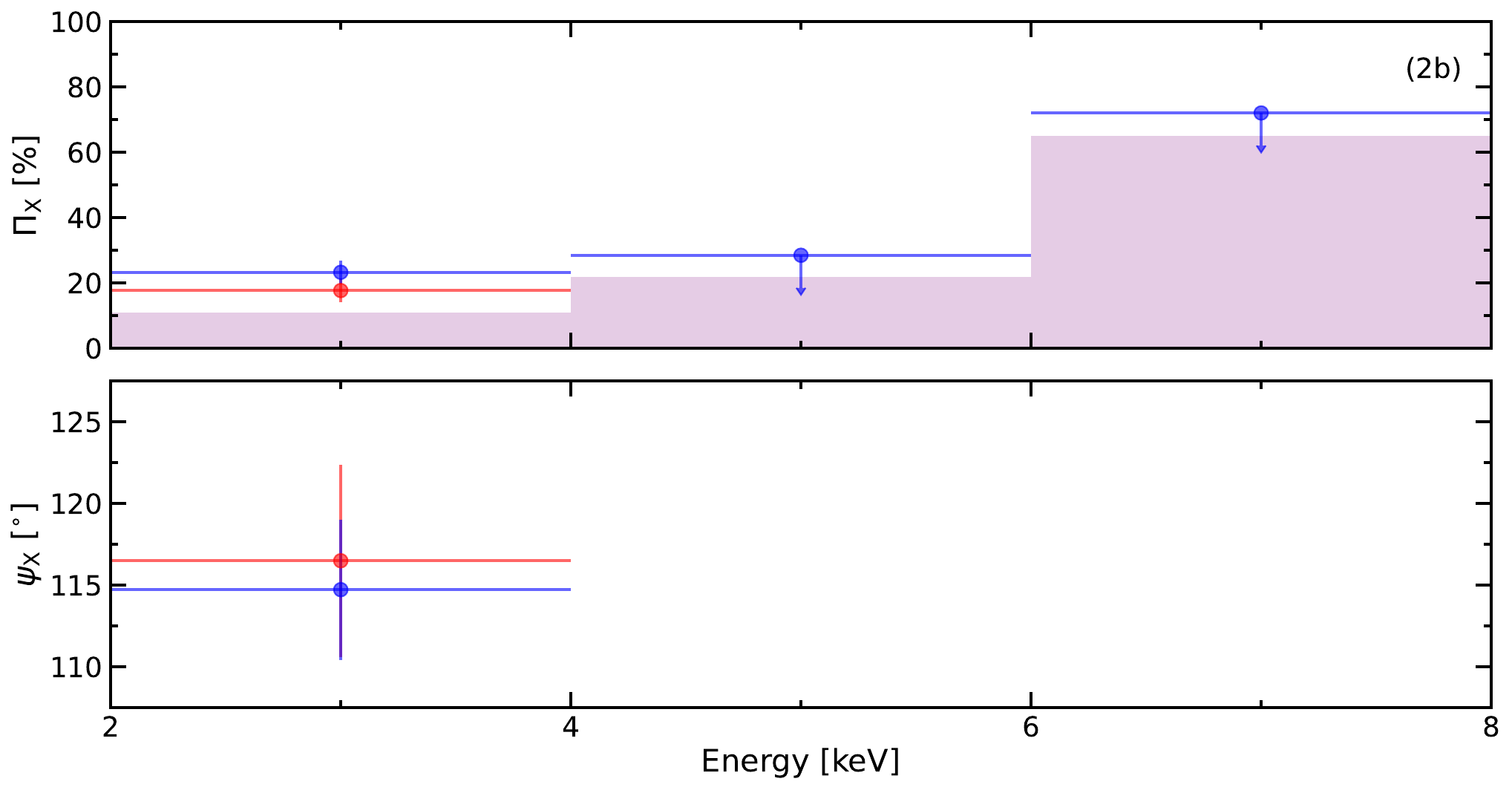}
    \includegraphics[angle=0, scale=0.25]{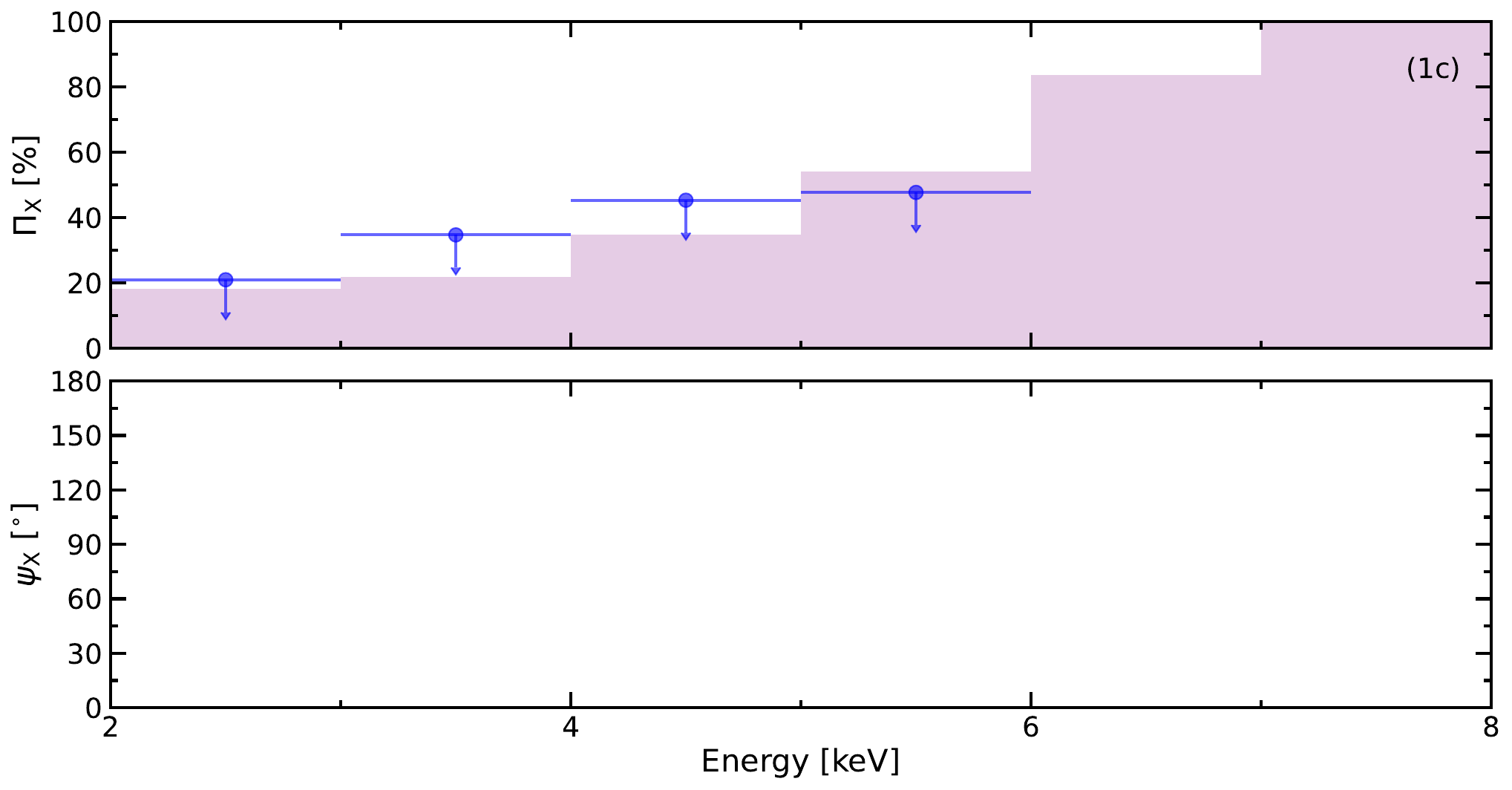}
    \includegraphics[angle=0, scale=0.25]{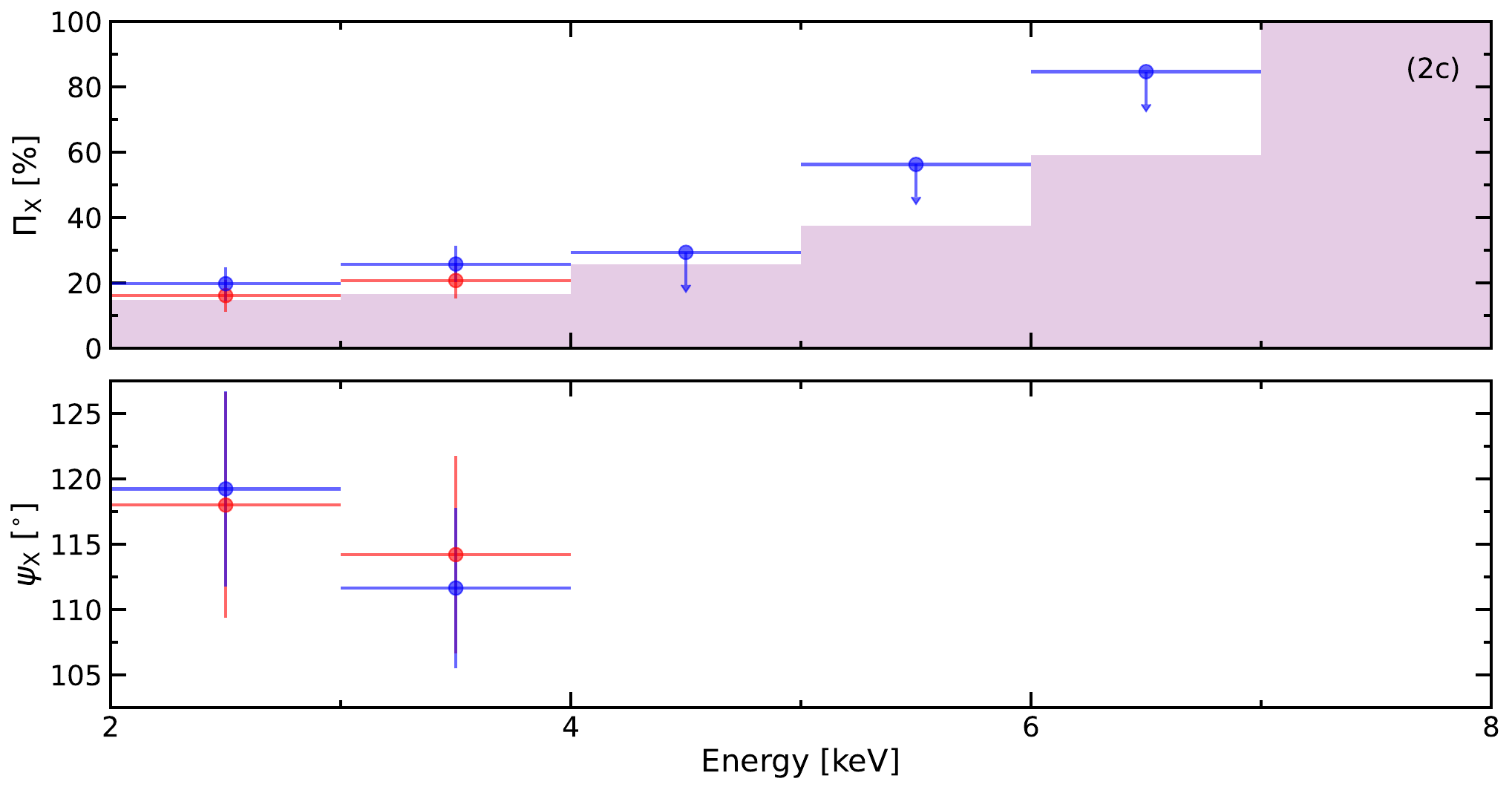}
    \caption{Results of energy-resolved polarization analysis for the first ({\it panels} (1)) and second ({\it panels} (2)) IXPE observations of H 1426+428 in different energy bins: {\it panels} (a) for 3 keV bin$^{-1}$, {\it panels} (b) for 2 keV bin$^{-1}$, and {\it panels} (c) for 1 keV bin$^{-1}$, respectively. In each panel, the red points represent the results estimated via \texttt{PCUBE} algorithm in \texttt{ixpeobssim}, the blue points represent the results estimated via spectropolarimetric fits in \texttt{Xspec}, and the violet shaded areas represent the values of MDP$_{99}$ for different energy bins, respectively. Particularly, if the best-fit values of $\Pi_{\rm X}$ obtained from spectropolarimetric fits are lower than the associated values of MDP$_{99}$, upper limits of $\Pi_{\rm X}$ at 99\% C.L. will be provided, marked by blue points with down-arrows.}
    \label{fig_ebin}
\end{figure*}

Additionally, we conducted an analysis of the energy-dependent polarization for both IXPE observations. Notably, no discernible X-ray polarization of H 1426+428 is observed across any energy bin during the first IXPE observation, while the significant polarization observed during the second IXPE observation was predominantly driven by the emissions below 5 keV, as illustrated in Figure \ref{fig_ebin}. As depicted in the broadband SED (Figure \ref{fig_sed}), the synchrotron peak during the second IXPE observation is anticipated to exceed 5 keV. Given that the high-energy electron radiation region is situated closer to the particle acceleration site \citep[e.g.,][]{2022Natur.611..677L,2022ApJ...938L...7D,2023ApJ...953L..28M} and is predominantly governed by a few zones \citep{2019ApJ...885...76P}, it is theoretically predicted that higher polarization degrees will be observed in higher energy bands. Despite the signal-to-noise ratios above 5 keV not being excessively low (see Figure \ref{fig_specpol}) and the X-rays from H 1426+428 exhibiting a flat spectrum during the second IXPE observation, the effective area of IXPE decreases significantly at energies exceeding 6 keV (see Figure 5.1 in the IXPE User Guide-Observatory \footnote{\url{https://heasarc.gsfc.nasa.gov/docs/ixpe/analysis/}}). Therefore, non-detection of polarization above 5 keV is likely attributed to the limited photon statistics. This inference is further supported by the high values of MDP$_{99}$ in high-energy bins. This scenario is different from that of the LSP BL Lacertae, where the IXPE observation energy band encompasses the synchrotron-SSC transition region, leading to the absence of polarization above 4 keV \citep{2023ApJ...948L..25P}. Additionally, considering the higher values of MDP$_{99}$ across all energy bins for the first IXPE observation compared to the second one, along with the high upper limit value of $\Pi_{\rm X}<19.5\%$ in the 2--8 keV band for the first IXPE observation, as illustrated Figures \ref{fig_qu} and \ref{fig_ebin}, we are unable to conclusively confirm that the magnetic fields in the plasma responsible to the X-ray emission during the first IXPE observation were entirely disordered, resulting in no significant detection of X-ray polarization. In other words, the non-detection of polarization during the first observation may also be attributed to the limited photon counts.

Most of the HSPs observed by IXPE to date share a common characteristic: their $\Pi$ across a broad energy range exhibits chromatic behavior, as found in Mrk 501 \citep{2022Natur.611..677L,2024ApJ...970L..22H}, Mrk 421 \citep{2022ApJ...938L...7D,2024A&A...681A..12K}, PG 1553+113 \citep{2023ApJ...953L..28M}, 1ES 0229+200 \citep{2023ApJ...959...61E}, and PKS 2155-304 \citep{2024A&A...689A.119K}. We analyze the long-term optical polarimetry data of H 1426+428 from the Steward Observatory spectropolarimetric monitoring project\footnote{\url{https://james.as.arizona.edu/~psmith/Fermi/}} \citep{2009arXiv0912.3621S}, finding that the average optical polarization degree ($\Pi_{\rm O}$) is merely about $\Pi_{\rm O}=1.1\%\pm0.1\%$ between 2008 November 30 and 2012 March 26, with the highest value reaching $\Pi_{\rm O}=2.23\%\pm0.05\%$. Recently, \citet{2025arXiv250412410B} reported that the optical polarization measurements, conducted simultaneously with IXPE observations, indicate a polarization degree of $\Pi_{\rm O}<3\%$. These are significantly lower than that in the 2--8 keV band of $\Pi_{\rm X}=20.6\%\pm2.9\%$. The 1991 Very Long Baseline Interferometry observation at 5 GHz revealed no detected polarized flux, with a jet position angle of $20\degr$ for H 1426+428 \citep{1996ApJ...460..174K}. However, the 8.4 GHz VLBA images observed in 2001 and 2003 indicate that the jet position angle is $155\degr$ \citep{2008ApJ...678...64P}. It seems likely that H 1426+428 also exhibits the chromatic behavior of $\Pi$, as observed in other HSPs \citep[e.g.,][]{2022Natur.611..677L,2022ApJ...938L...7D,2023ApJ...953L..28M,2023ApJ...959...61E,2024A&A...689A.119K}. Considering the upper limit of $\Pi_{\rm O}<3\%$ reported by \citet{2025arXiv250412410B}, the ratio of X-ray to optical polarization degrees (i.e., $\Pi_{\rm X}/\Pi_{\rm O}$) for H 1426+428 is approximately 7 during the second IXPE observation. A similarly high ratio was also observed in another EHSP, 1ES 0229+200 \citep{2023ApJ...959...61E}. Clearly, the derived polarization angles in the X-ray (this work) and optical \citep{2025arXiv250412410B} bands are not aligned with the jet position angle, as previously observed in Mrk 421 \citep{2022ApJ...938L...7D}. The discrepancy between the radio jet position angle and the multiwavelength polarization angle may imply that the jet undergoes bending between the emission regions of the high- and low-energy electrons \citep{2008ASPC..386..437M}, or the emission regions of electrons at X-ray and radio energies are spatially separated \citep{2009MNRAS.397..985G}, as discussed in \citet{2022ApJ...938L...7D}.

The synchrotron cooling time ($t_{\rm s}$) of relativistic electrons can be estimated as $t_{\rm s}\sim7.75\times10^{8}B^{-2}\gamma^{-1}$ \citep[in seconds;][]{1999MNRAS.306..551C}, where $B$ is the magnetic field strength and $\gamma$ is the Lorentz factor of the electrons. Based on the synchrotron radiation frequency $\nu_{\rm s}=3.7\times10^{6}B\gamma^{2}\frac{\delta}{1+z}$ \citep{1998ApJ...509..608T}, the synchrotron cooling time is further evaluated as $t_{\rm s}\sim1.5\times10^{12}B^{-3/2}\nu_{\rm s}^{-1/2}\delta^{1/2}(1+z)^{-1/2}$ (in seconds), where $\delta$ is the beaming factor. Using the derived parameters $B=0.1~{\rm G}$ and $\delta=8.5$ from broadband SED fits, which are also representative values for TeV-emitting BL Lacs \citep{2012ApJ...752..157Z}, the cooling time of electrons responsible for the X-ray emission in the 2--8 keV band is estimated to be approximately 2 days. Given that the time interval between the two IXPE observations spans approximately 36 days, the electron population responsible for the X-ray emission during the first IXPE observation is unlikely to contribute to the X-ray emission during the second IXPE observation unless it has been re-accelerated. The X-ray observations of H 1426+428 during the second IXPE observation reveal a higher flux state with a harder spectrum compared to that during the first IXPE observation, as shown in Figures \ref{fig_flux-gamma} and \ref{fig_sed}. This {\it harder-when-brighter} spectral behavior, commonly observed in the X-ray emission of HSPs \citep[e.g.,][]{2009A&A...501..879T,2022ApJ...938L...7D,2024ApJ...963L..41H,2024ApJ...970L..22H}, is widely interpreted as the injection of fresh high-energy electrons by a shock \citep{1998A&A...333..452K,2022ApJ...938L...7D}. Therefore, we suppose that the plasma responsible for the X-ray emission during the first IXPE observation propagates downstream and subsequently encounters a shock, leading to electron re-acceleration, or alternatively, a moving shock propagates through the plasma, as discussed in the context of the blazar S4 0954+65 \citep{2025A&A...695A..99K}, resulting in the injection of fresh high-energy electrons. Due to the plasma processes \citep{2018MNRAS.480.2872T} or compression \citep{1985ApJ...298..301H}, partially ordered magnetic fields can be generated near the shock \citep{2022ApJ...938L...7D}. As a result, during the second IXPE observation, the X-rays are characterized by a higher flux state, a harder spectrum, and significant polarization. Furthermore, shock acceleration of particles naturally explains the chromatic behavior of the polarization degree observed in these HSPs by IXPE \citep[e.g.,][]{2022Natur.611..677L,2022ApJ...938L...7D}, including H 1426+428.

\begin{figure*}
    \centering
    \includegraphics[angle=0, scale=0.35]{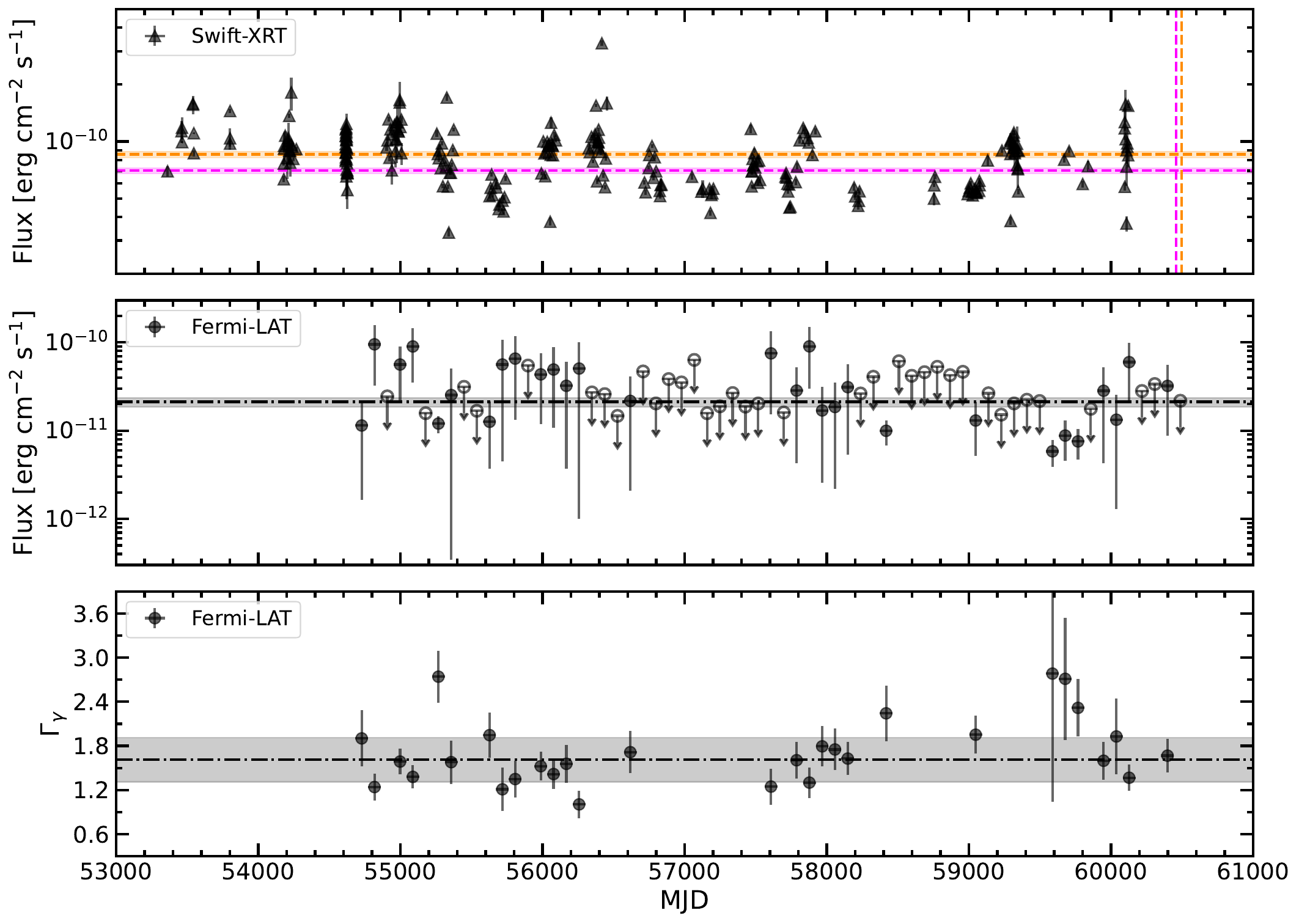}
    \caption{Long-term light curves of H 1426+428 derived from Swift-XRT observations ({\it top panel}) and Fermi-LAT observations ({\it middle panel}), along with the corresponding $\Gamma_{\gamma}$ values for each detection point in the $\gamma$-ray light curve ({\it bottom panel}). The X-ray light curve in the 0.3--10 keV band is obtained from the long-term Swift-XRT monitoring program of Fermi-LAT sources of interest \citep{2013ApJS..207...28S}. The magenta and orange vertical dashed lines indicate the observation dates of Swift-XRT that are quasi-simultaneous with the first and second IXPE observations of H 1426+428, respectively. The horizontal lines correspond to the flux values measured by Swift-XRT in the 0.3--10 keV band. The $\gamma$-ray light curve in the 0.1--1000 GeV band derived from Fermi-LAT observations is binned in 90-day intervals. Open circles with down-arrows denote upper limit values when test statistic (TS) value < 9 for a given time bin. The dash-dot lines in the middle and bottom panels represent the average flux and photon spectral index values derived with the $\sim$ 16-year Fermi-LAT observations. The shaded horizontal regions denote the 1$\sigma$ uncertainty intervals corresponding to the aforementioned average values.}
    \label{fig_lc}
\end{figure*}

\begin{figure*}
    \centering
    \includegraphics[angle=0, scale=0.45]{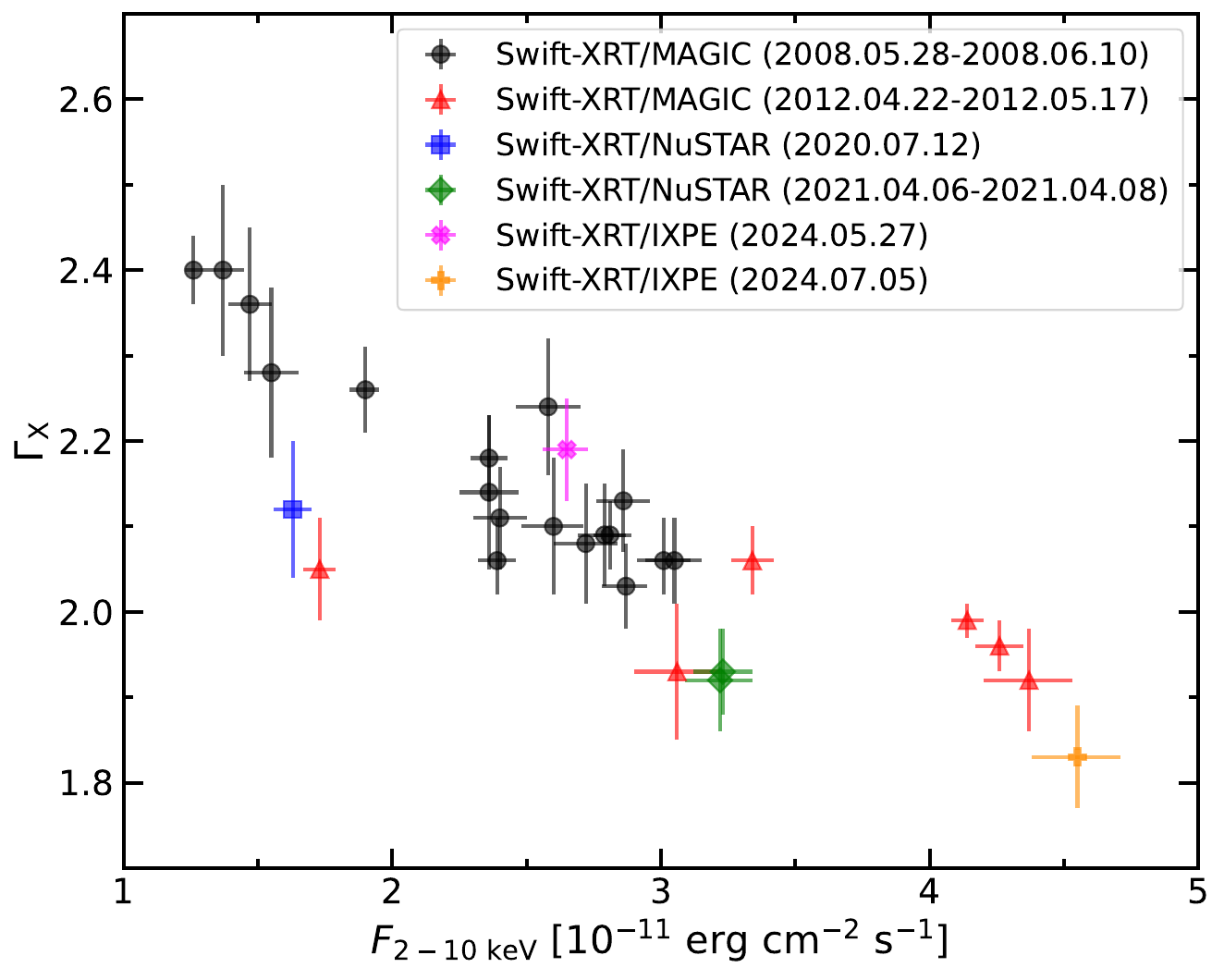}
    \caption{$\Gamma_{\rm X}$ as a function of $F_{\rm 2-10~keV}$ for H 1426+428, where the Swift-XRT data are identical to those presented in Table \ref{tab_xrt}. The black and red symbols represent the results quasi-simultaneously observed during the two MAGIC observations. The blue and green symbols correspond to the results quasi-simultaneously observed during the two NuSTAR observations. The magenta and orange symbols denote the results quasi-simultaneously observed during the two IXPE observations.}
    \label{fig_flux-gamma}
\end{figure*}

As displayed in Figure \ref{fig_lc}, the X-ray fluxes exhibit significant variability, whereas the GeV $\gamma$-ray fluxes remain relatively stable. Through analysis of the Swift-XRT observational data contemporaneously acquired with the two MAGIC observations, it appears that the TeV detection is associated with a high X-ray flux state (as illustrated in Figure \ref{fig_sed}). The broadband SEDs of TeV BL Lacs are typically explained by the one-zone leptonic model \citep[e.g.,][]{1998ApJ...509..608T,2012ApJ...752..157Z,2014MNRAS.439.2933Y}. As depicted in the Figure 2 of \citet{1998ApJ...509..608T}, the X-rays and TeV $\gamma$-rays are likely produced by high-energy electrons through synchrotron and SSC processes, respectively, whereas the GeV $\gamma$-rays are predominantly generated by low-energy electrons through the SSC process.

\begin{acknowledgments}

This work reports observations obtained with the Imaging X-ray Polarimetry Explorer (IXPE), a joint US (NASA) and Italian (ASI) mission, led by Marshall Space Flight Center (MSFC). The research uses data products provided by the IXPE Science Operations Center (MSFC), using algorithms developed by the IXPE Collaboration, and distributed by the High-Energy Astrophysics Science Archive Research Center (HEASARC). Data from the Steward Observatory spectropolarimetric monitoring project are used. This program is supported by Fermi Guest Investigator grants NNX08AW56G, NNX09AU10G, NNX12AO93G, and NNX15AU81G.

We thank the referee for the valuable suggestions that improved the manuscript. This work is supported by the National Key R\&D Program of China (grant 2023YFE0117200) and the National Natural Science Foundation of China (grants 12203022, 12022305, 11973050, and 12373041).

\end{acknowledgments}

\clearpage 

\bibliography{reference}
\bibliographystyle{aasjournal}

\clearpage

\begin{table*}
    \begin{center}
    \caption{Spectral and Spectropolarimetric Analysis Results for H 1426+428}
    \label{tab_specpol}
    \begin{tabular}{ccccc}
    \hline
    \hline
    Component & Parameter & Unit & \multicolumn{2}{c}{Value} \\
    \cmidrule(r){4-5}
    & & & 03007201 & 03007301 \\
    \hline
    \multicolumn{5}{c}{Model = \texttt{CONSTANT}$\times$\texttt{TBABS}$\times$\texttt{POWERLAW}} \\
    \hline
    \texttt{CONSTANT} & $C_{\rm DU1}$ & & 1.0 (fixed) & 1.0 (fixed) \\
    & $C_{\rm DU2}$ & & $1.025^{+0.022}_{-0.021}$ & $1.018\pm0.017$ \\
    & $C_{\rm DU3}$ & & $0.984\pm0.021$ & $0.988\pm0.017$ \\
    & $C_{\rm XRT}$ & & $1.122^{+0.055}_{-0.054}$ & $0.989^{+0.053}_{-0.052}$ \\
    \texttt{TBABS} & $N_{\rm H}$ & $10^{20}$ cm$^{-2}$ & $9.65^{+1.39}_{-1.31}$ & $5.48^{+1.60}_{-1.45}$ \\
    \texttt{POWERLAW} & $\Gamma_{\rm X}$ & & $2.29\pm0.04$ & $2.00\pm0.03$ \\
    & $N_{0}$ & $10^{-2}~{\rm ph~cm^{-2}~s^{-1}~keV^{-1}}$ & $1.24\pm0.05$ & $1.44\pm0.05$ \\
    Fit Statistic & $\chi^{2}$/dof & & 258/257 & 255/286 \\
    Flux & $F_{\rm 2-8~keV}$ & $10^{-11}~{\rm erg~cm^{-2}~s^{-1}}$ & $1.85\pm0.03$ & $3.21\pm0.04$ \\
    \hline
    \multicolumn{5}{c}{Model = \texttt{CONSTANT}$\times$\texttt{TBABS}$\times$\texttt{POLCONST}$\times$\texttt{POWERLAW}} \\
    \hline
    \texttt{POLCONST} & $\Pi_{\rm X}$ & \% & $<19.5$ & $20.6\pm2.9$ \\
    & $\psi_{\rm X}$ & $\degr$ & \nodata & $116.1\pm4.1$ \\
    Fit Statistic & $\chi^{2}$/dof & & 370/387 & 351/425 \\
    \hline
    \hline
    \end{tabular}
    \end{center}
\end{table*}

\clearpage

\clearpage

\appendix
\section{Observations in Multiwavelength}

\subsection{Swift-XRT}\label{sec_xrt}

XRT \citep[i.e., the X-ray Telescope;][]{2005SSRv..120..165B} on board the Neil Gehrels Swift Observatory \citep[Swift;][]{2004ApJ...611.1005G} carried out multiple periodic observations of H 1426+428. In this work, we focus on six specific observation periods: two concurrent with IXPE observations, two concurrent with NuSTAR observations, and two concurrent with MAGIC observations. Detailed information regarding the data analysis can be found below.

The Swift-XRT observations were performed in Photon Counting (PC) or Window Timing (WT) readout modes. The data are processed using the XRT Data Analysis Software (XRTDAS, v.3.7.0) within the \texttt{HEASoft} package. Calibration of the observational event files is achieved using calibration files from the Swift-XRT CALDB (v.20240522) through the \texttt{xrtpipeline} task. To eliminate the pile-up effects in the observations during PC readout mode, source photons are extracted from an annular region centered on the brightest pixel, with inner and outer radii of $6^{\prime\prime}$--$15^{\prime\prime}$ and $47^{\prime\prime}$, respectively. Since there is no pile-up effect in the observations with WT readout mode, source photons are extracted from a circular region with a radius of $47^{\prime\prime}$ also centered on the brightest pixel. Background photons are extracted from a homocentric annular region with inner and outer radii of $71^{\prime\prime}$ and $142^{\prime\prime}$, respectively. The source and background spectra are generated using the \texttt{xselect} task. The ancillary response files (ARFs) are created using the cumulative exposure maps through the \texttt{xrtmkarf} task, accounting for PSF losses and CCD defects.

The Swift-XRT observations conducted on 2024 May 27 and 2024 July 5 were performed in PC readout mode, which are simultaneous with the two IXPE observations. Similar to the IXPE data reduction process, the Swift-XRT spectra are grouped to ensure a minimum of 20 counts per energy bin. The procedure for the fits of the joint IXPE and Swift-XRT spectra is detailed in Section \ref{sec_ixpe}. The joint spectra are presented in Figure \ref{fig_spec_ixpe+xrt}.

The Swift-XRT spectra, which are quasi-simultaneously observed with the NuSTAR observations (details can be found in Section \ref{sec_nustar}), are jointly fitted with the NuSTAR spectra using an absorbed PL model within \texttt{Xspec}. This approach is consistent with the procedures detailed in Section \ref{sec_ixpe}. We define the energy ranges as 3--30 keV for the NuSTAR spectra and 0.3--10 keV for the Swift-XRT spectra, respectively, thereby obtaining a joint spectral range of 0.3--30 keV. Given the brief time interval, we merge the event files from the two Swift-XRT observations that are quasi-simultaneous with the second NuSTAR observation to generate a composite average spectrum. During the spectral fits, the parameters $N_{\rm H}$, $\Gamma_{\rm X}$, and $N_{0}$ are allowed to vary freely.

For the 24 Swift-XRT observations that are quasi-simultaneous with two MAGIC observations, the spectra are also fitted using an absorbed PL model. However, the $N_{\rm H}$ values cannot be well constrained compared with the situation that the Swift-XRT spectra are jointly fitted with the IXPE or NuSTAR spectra. It is evident that the $N_{\rm H}$ constraint values derived from joint spectral fits are significantly higher than the Galactic value of H 1426+428 \citep[i.e., $N_{\rm H}=0.95\times10^{20}~{\rm cm^{-2}}$;][]{2016A&A...594A.116H}, as detailed in Tables \ref{tab_specpol} and \ref{tab_xrt+nustar}. Therefore, $N_{\rm H}$ is fixed at a statistically determined value of $N_{\rm H}=5.20\times10^{20}~{\rm cm^{-2}}$, corresponding to the average value reported in \citet{2013ApJS..207...28S}, which is obtained through the long-term Swift-XRT monitoring program for Fermi-LAT sources of interest. In this process, $\Gamma_{\rm X}$ and $N_{0}$ are treated as free parameters. The best-fit parameters are presented in Table \ref{tab_xrt}.

For comparison, we also independently analyze the Swift-XRT spectra that are simultaneous with the NuSTAR and IXPE observations using an absorbed PL model, where $N_{\rm H}$ is fixed at the values derived from the combined spectra fits. The results are also presented in Table \ref{tab_xrt}.

\subsection{NuSTAR}\label{sec_nustar}

NuSTAR \citep[i.e., the Nuclear Spectroscopic Telescope Array;][]{2013ApJ...770..103H}, equipped with two multilayer-coated telescopes, FPMA and FPMB, conducted two observations of H 1426+428 on 2020 July 12 (PI: Costamante) and on 2021 April 7 (PI: Harrison). In this study, we analyze the data from these two observations. The raw data are processed using the \texttt{nupipeline} task within the NuSTAR Data Analysis Software (NuSTARDAS, v.2.1.4). The calibration files from NuSTAR CALDB (v.20240826) are used to produce calibrated and cleaned event files. Source photons are extracted from a circular region with a radius of $90^{\prime\prime}$ centered on the centroid of the X-ray emission, while background photons are extracted from a nearby circular region with a radius of $120^{\prime\prime}$. The source and background spectra, along with the RMFs and ARFs, are generated using the \texttt{nuproducet} task.The best-fit parameters of the joint Swift-XRT and NuSTAR spectral fits are summarized in Table \ref{tab_xrt+nustar} and the joint spectra are illustrated in Figure \ref{fig_spec_xrt+nustar}.

\subsection{Fermi-LAT}\label{sec_lat}

H 1426+428 is reported to be associated with the $\gamma$-ray source 4FGL J1428.5+4240 in the 4FGL-DR4 catalog \citep{2022ApJS..260...53A,2023arXiv230712546B}. We download the PASS 8 data covering from 2008 August 4 to 2024 November 4 (MJD 54682--60618) for the energy range of 0.1--1000 GeV within a $15\degr$ region of interest (ROI) centered on the HST pointing position of H 1426+428 from the Fermi Science Support Center \footnote{\url{https://fermi.gsfc.nasa.gov/cgi-bin/ssc/LAT/LATDataQuery.cgi}}. The publicly available software \texttt{fermitools} (v.2.2.0) and \texttt{Fermipy} \citep[v.1.1;][]{2017ICRC...35..824W} are employed in the data analysis. The event class "SOURCE" (evclass = 128) and event type "FRONT+BACK" (evtype = 3) for the binned likelihood analysis are utilized based on LAT data selection recommendations \footnote{\url{https://fermi.gsfc.nasa.gov/ssc/data/analysis/documentation/Cicerone/Cicerone_Data_Exploration/Data_preparation.html}}. The filter expression "(DATA\_QUAL>0)\&\&(LAT\_CONFIG==1)" and the instrument response function of $\rm P8R3\_SOURCE\_V3$ are adopted and the maximum zenith angle of 90$\degr$ is set aiming to eliminate the contamination of $\gamma$-ray emission from the earth limb. All the sources within the 4FGL-DR4 catalog are added to the source model as well as two background model, namely the isotropic emission ("$\rm iso\_P8R3\_SOURCE\_V3\_V1.txt$") and the diffuse galactic interstellar emission ("$\rm gll\_iem\_v07.fits$"). During the data processing, the parameters of those $\gamma$-ray sources within $6.0\degr$ centered on H 1426+428 and the normalization parameters of two background models are kept free.

The TS value is used to evaluate the significance of a $\gamma$-ray source signal. The TS value is calculated as
\begin{equation}
{\rm TS}=2\left(\log\mathcal{L}_{\rm Src}-\log\mathcal{L}_{\rm Null}\right),
\end{equation}
where $\mathcal{L}_{\rm Src}$ and $\mathcal{L}_{\rm Null}$ are the likelihood values of the sky region with and without the target source, respectively. We conduct a residual TS map for the sky region around H 1426+428. The obtained maximum TS value is $\sim4.08$, which indicates that no new source is detected.

H 1426+428 is identified as a point source with a PL spectrum in 4FGL-DR4 catalog \citep{2022ApJS..260...53A,2023arXiv230712546B}. We also use a PL function to fit the spectrum, which is defined as
\begin{equation}
\frac{dN}{dE}=N_0\times\left(\frac{E}{E_0}\right)^{-\Gamma_\gamma}.
\end{equation}
By analyzing the $\sim$ 16-year Fermi-LAT observational data of H 1426+428, we obtain an average flux of $(2.12\pm0.23)\times10^{-11}~{\rm erg~cm^{-2}~s^{-1}}$ in the 0.1--1000 GeV band, with a TS value of $\sim$ 1153.10 and a photon spectral index of $\Gamma_{\gamma}=1.61\pm0.03$.

The long-term light curve with equal time bin of 90 days in the 0.1--1000 GeV band is conducted to investigate the $\gamma$-ray variability of H 1426+428, as shown in Figure \ref{fig_lc}. The variability of the long-term $\gamma$-ray light curve is assessed by estimating the $P_{\rm Null}$ through $\chi^{2}$ test, similar to the time-resolved analysis of X-ray polarization described in Section \ref{sec_results}. The $\chi^{2}$ test reveals that no significant variability of H 1426+428 in the GeV $\gamma$-ray band is observed during the $\sim$ 16-year Fermi-LAT observation, with $P_{\rm Null}>50\%$.

\subsection{Broadband SED}\label{sec_sed}

Using the data from NED, the MAGIC observations \citep{2009arXiv0907.0959L,2020ApJS..247...16A}, and the analysis results of Swift-XRT and Fermi-LAT observations in this study, we construct the broadband SED of H 1426+428, as shown in Figure \ref{fig_sed}. It is worth noting that the X-ray flux levels during the two IXPE observations are comparable to those during the two MAGIC observations, respectively. Given the absence of significant variability in the long-term light curve of GeV $\gamma$-rays, the $\sim$ 16-year average spectrum of the Fermi-LAT observations is presented in Figure \ref{fig_sed}, where an upper limit is presented for that energy bin if TS < 9.

\clearpage

\renewcommand{\thetable}{A.1}
\begin{table*}
    \begin{center}
    \caption{Analysis Results of the Swift-XRT Spectral Fits for H 1426+428}
    \label{tab_xrt}
    \begin{tabular}{ccccccccc}
    \hline
    \hline
    OBSID & Date & Exposure & Mode & $N_{\rm H}$ & $\Gamma_{\rm X}$ & $N_{0}$$^{\rm d}$ & Flux$^{\rm e}$ & $\chi^{2}$/dof \\
    & & (s) & & ($10^{20}~{\rm cm^{-2}}$) & & & ($10^{-11}~{\rm erg~cm^{-2}~s^{-1}}$) & \\
    \hline
    00030375015 & 2008.05.28 & 2408 & PC & 5.20$^{\rm a}$ & $2.06^{+0.05}_{-0.04}$ & $1.02\pm0.03$ & $2.39\pm0.07$ & 59/55 \\
    00030375016 & 2008.05.29 & 1704 & PC & 5.20$^{\rm a}$ & $2.11\pm0.06$ & $1.10\pm0.05$ & $2.40\pm0.10$ &  38/28 \\
    00030375017 & 2008.05.29 & 1473 & PC & 5.20$^{\rm a}$ & $2.08\pm0.07$ & $1.19\pm0.05$ & $2.72\pm0.12$ &  37/27 \\
    00030375018 & 2008.05.30 & 1986 & PC & 5.20$^{\rm a}$ & $2.13\pm0.06$ & $1.34\pm0.05$ & $2.86\pm0.10$ & 36/39 \\
    00030375019 & 2008.05.30 & 3910 & PC & 5.20$^{\rm a}$ & $2.09\pm0.04$ & $1.24\pm0.03$ & $2.81\pm0.06$ & 76/88 \\
    00030375020 & 2008.05.31 & 2279 & PC & 5.20$^{\rm a}$ & $2.03\pm0.05$ & $1.18\pm0.04$ & $2.87^{+0.08}_{-0.09}$ & 39/53 \\
    00030375021 & 2008.06.01 & 2095 & PC & 5.20$^{\rm a}$ & $2.06\pm0.05$ & $1.29\pm0.05$ & $3.05^{+0.10}_{-0.11}$ & 63/42 \\
    00030375022 & 2008.06.01 & 1761 & PC & 5.20$^{\rm a}$ & $2.06^{+0.05}_{-0.04}$ & $1.28\pm0.04$ & $3.01\pm0.10$ & 51/45 \\
    00030375023 & 2008.06.02 & 1450 & PC & 5.20$^{\rm a}$ & $2.09\pm0.06$ & $1.24\pm0.05$ & $2.79\pm0.10$ & 38/37 \\
    00030375024 & 2008.06.03 & 1401 & PC & 5.20$^{\rm a}$ & $2.24\pm0.08$ & $1.42\pm0.07$ & $2.58\pm0.12$ & 16/21 \\
    00030375025 & 2008.06.03 & 1732 & PC & 5.20$^{\rm a}$ & $2.10\pm0.08$ & $1.17\pm0.05$ & $2.60^{+0.11}_{-0.12}$ & 19/25 \\
    00030375026 & 2008.06.04 & 1029 & PC & 5.20$^{\rm a}$ & $2.14\pm0.09$ & $1.13\pm0.05$ & $2.36\pm0.11$ & 30/23 \\
    00030375027 & 2008.06.04 & 3237 & PC & 5.20$^{\rm a}$ & $2.18\pm0.05$ & $1.19\pm0.04$ & $2.36\pm0.07$ & 40/55 \\
    00030375028 & 2008.06.06 & 3482 & PC & 5.20$^{\rm a}$ & $2.26\pm0.05$ & $1.07\pm0.03$ & $1.90^{+0.05}_{-0.06}$ & 58/55 \\
    00030375029 & 2008.06.07 & 975 & PC & 5.20$^{\rm a}$ & $2.36\pm0.09$ & $0.96\pm0.05$ & $1.47\pm0.08$ & 16/17 \\
    00030375030 & 2008.06.08 & 778 & PC & 5.20$^{\rm a}$ & $2.28\pm0.10$ & $0.91\pm0.06$ & $1.55\pm0.10$ & 3/10 \\
    00030375031 & 2008.06.09 & 1114 & PC & 5.20$^{\rm a}$ & $2.40\pm0.10$ & $0.96\pm0.05$ & $1.37\pm0.08$ & 9/14 \\
    00030375032 & 2008.06.09 & 3558 & PC & 5.20$^{\rm a}$ & $2.40\pm0.04$ & $0.88\pm0.02$ & $1.26\pm0.03$ & 83/83 \\
    00030375086 & 2012.04.22 & 344 & WT & 5.20$^{\rm a}$ & $1.92\pm0.06$ & $1.50\pm0.06$ & $4.37^{+0.16}_{-0.17}$ & 34/31 \\
    00030375087 & 2012.04.28 & 1010 & WT & 5.20$^{\rm a}$ & $1.96\pm0.03$ & $1.56\pm0.03$ & $4.26\pm0.09$ & 75/90 \\
    00030375088 & 2012.05.07 & 1155 & WT & 5.20$^{\rm a}$ & $2.05\pm0.06$ & $0.73\pm0.03$ & $1.73\pm0.06$ & 37/37 \\
    00030375089 & 2012.05.10 & 974 & WT & 5.20$^{\rm a}$ & $2.06\pm0.04$ & $1.42\pm0.04$ & $3.34\pm0.08$ & 75/75 \\
    00030375090 & 2012.05.13 & 299 & PC & 5.20$^{\rm a}$ & $1.93\pm0.08$ & $1.07\pm0.06$ & $3.06\pm0.16$ & 13/16 \\
    00030375091 & 2012.05.17 & 2029 & WT & 5.20$^{\rm a}$ & $1.99\pm0.02$ & $1.57\pm0.02$ & $4.14\pm0.06$ & 145/167 \\
    00089122001 & 2020.07.12 & 1933 & PC & 7.36$^{\rm b}$ & $2.12\pm0.08$ & $0.76\pm0.03$ & $1.63\pm0.07$ & 30/27 \\
    00030375203 & 2021.04.06 & 1652 & PC & 4.48$^{\rm b}$ & $1.92\pm0.06$ & $1.11\pm0.05$ & $3.22^{+0.12}_{-0.13}$ & 28/34 \\
    00030375204 & 2021.04.08 & 1724 & PC & 4.48$^{\rm b}$ & $1.93\pm0.05$ & $1.13\pm0.04$ & $3.23\pm0.11$ & 34/42 \\
    00035020021 & 2024.05.27 & 1619 & PC & 9.65$^{\rm c}$ & $2.19\pm0.06$ & $1.36\pm0.05$ & $2.65^{+0.08}_{-0.09}$ & 58/46 \\
    00030375219 & 2024.07.05 & 1885 & PC & 5.48$^{\rm c}$ & $1.83\pm0.06$ & $1.37\pm0.06$ & $4.55^{+0.16}_{-0.17}$ & 44/37 \\
    \hline
    \hline
    \end{tabular}
    \end{center}
    \tablenotetext{\rm a}{$N_{\rm H}$ is fixed at the average value proveded in \citet{2013ApJS..207...28S}.}
    \tablenotetext{\rm b}{$N_{\rm H}$ is fixed at the best-fit value obtained from the combined spectra fits with quasi-simultaneous NuSTAR observations.}
    \tablenotetext{\rm c}{$N_{\rm H}$ is fixed at the best-fit value obtained from the combined spectra fits with quasi-simultaneous IXPE observations.}
    \tablenotetext{\rm d}{$N_{0}$ is in units of $10^{-2}~{\rm ph~cm^{-2}~s^{-1}~keV^{-1}}$.}
    \tablenotetext{\rm e}{The fluxes are estimated in the range of 2--10 keV.}
\end{table*}

\renewcommand{\thetable}{A.2}
\begin{table*}
    \begin{center}
    \caption{Analysis Results of the Joint Swift-XRT and NuSTAR Spectral Fits for H 1426+428}
    \label{tab_xrt+nustar}
    \begin{tabular}{cccccccccc}
    \hline
    \hline
    Date$^{\rm a}$ & Exposure$^{\rm b}$ & $C_{\rm XRT}$$^{\rm c}$ & $C_{\rm FPMA}$ & $C_{\rm FPMB}$ & $N_{\rm H}$ & $\Gamma_{\rm X}$ & $N_{0}$$^{\rm d}$ & Flux$^{\rm e}$ & $\chi^{2}$/dof \\
    & (s) & & & & ($10^{20}~{\rm cm^{-2}}$) & & & & \\
    \hline
    2020.07.12 & 54020 & 1.0 & $1.500^{+0.104}_{-0.094}$ & $1.488^{+0.104}_{-0.094}$ & $7.36^{+1.81}_{-1.68}$ & $2.29\pm0.01$ & $0.78\pm0.05$ & $1.42\pm0.06$ & 707/675 \\
    2021.04.07 & 27875 & 1.0 & $1.405^{+0.052}_{-0.050}$ & $1.394^{+0.052}_{-0.049}$ & $4.48^{+0.77}_{-0.72}$ & $2.01\pm0.01$ & $1.08\pm0.04$ & $3.04^{+0.07}_{-0.08}$ & 757/783 \\
    \hline
    \hline
    \end{tabular}
    \end{center}
    \tablenotetext{\rm a}{The start date of the NuSTAR observation.}
    \tablenotetext{\rm b}{The average exposure between FPMA and FPMB.}
    \tablenotetext{\rm c}{The scaling factor of Swift-XRT is fixed at 1.}
    \tablenotetext{\rm d}{$N_{0}$ is in units of $10^{-2}~{\rm ph~cm^{-2}~s^{-1}~keV^{-1}}$.}
    \tablenotetext{\rm e}{The fluxes are estimated in the range of 2--30 keV and in units of $10^{-11}~{\rm erg~cm^{-2}~s^{-1}}$.}
\end{table*}

\clearpage

\renewcommand{\thefigure}{A.1}
\begin{figure*}
    \centering
    \includegraphics[angle=0, scale=0.35]{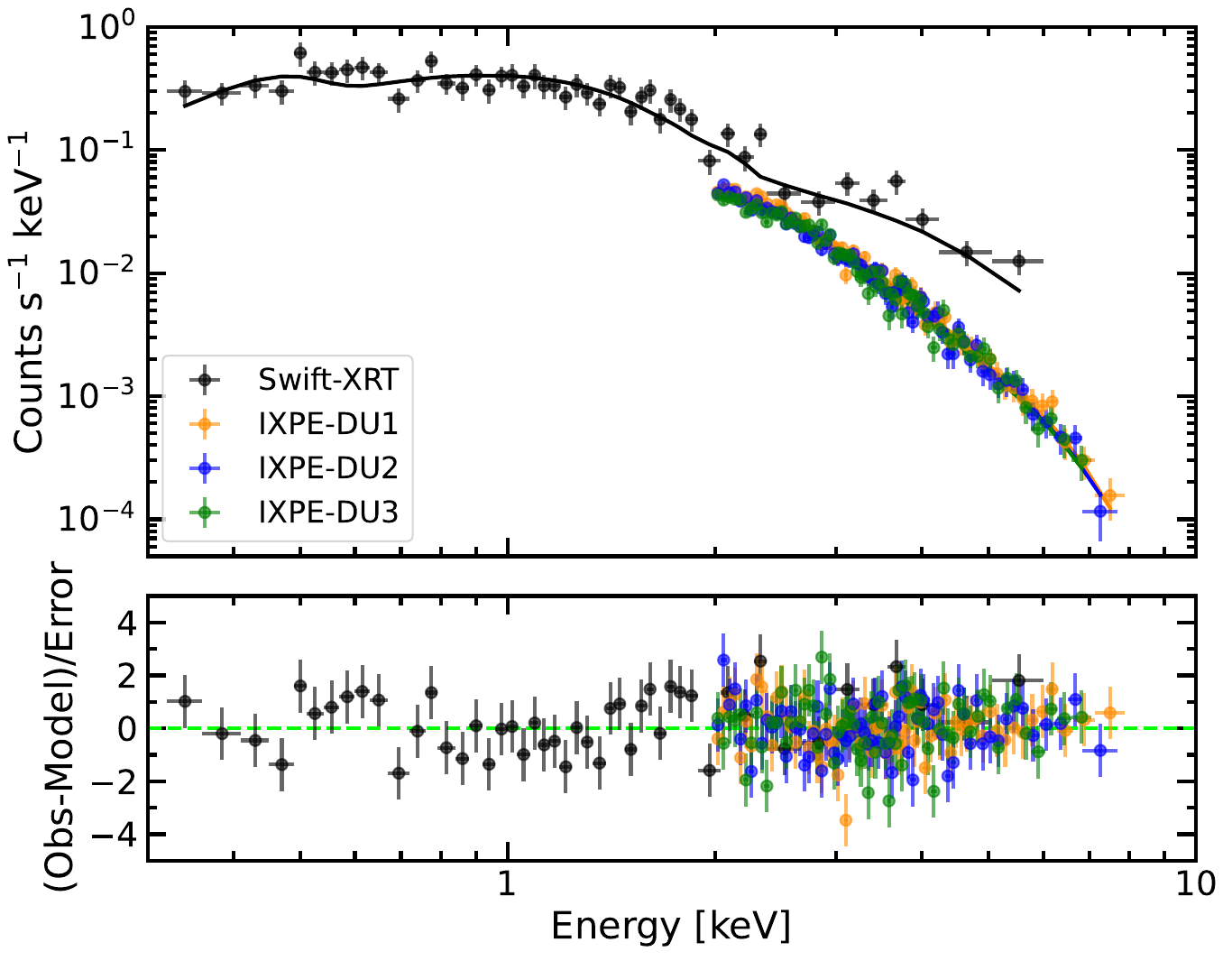}
    \includegraphics[angle=0, scale=0.35]{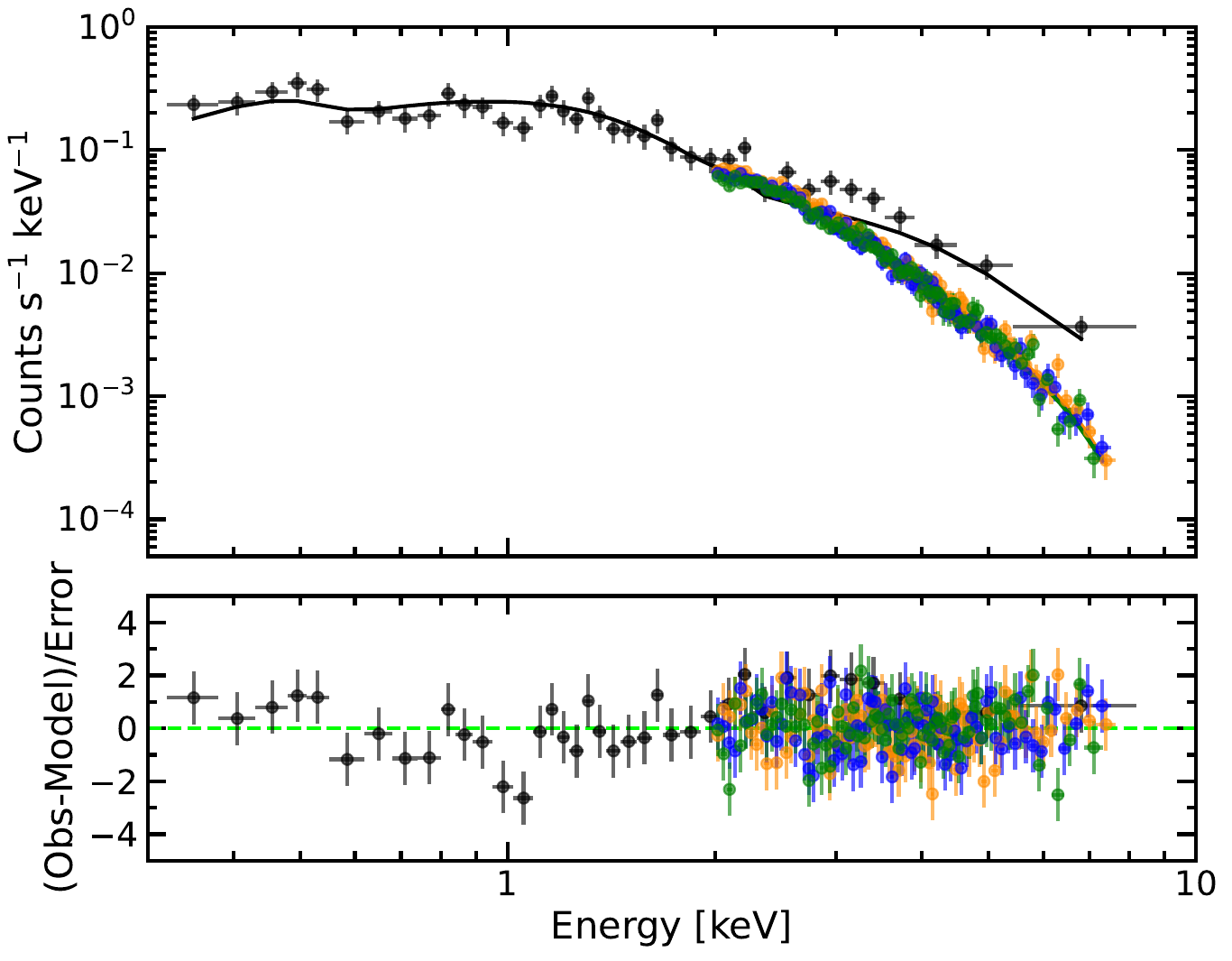}
    \caption{Results of the joint spectral fits using the IXPE and Swift-XRT spectra of H 1426+428, corresponding to the first first ({\it left panel}) and second ({\it right panel}) IXPE observations, respectively.}
    \label{fig_spec_ixpe+xrt}
\end{figure*}

\renewcommand{\thefigure}{A.2}
\begin{figure*}
    \centering
    \includegraphics[angle=0, scale=0.35]{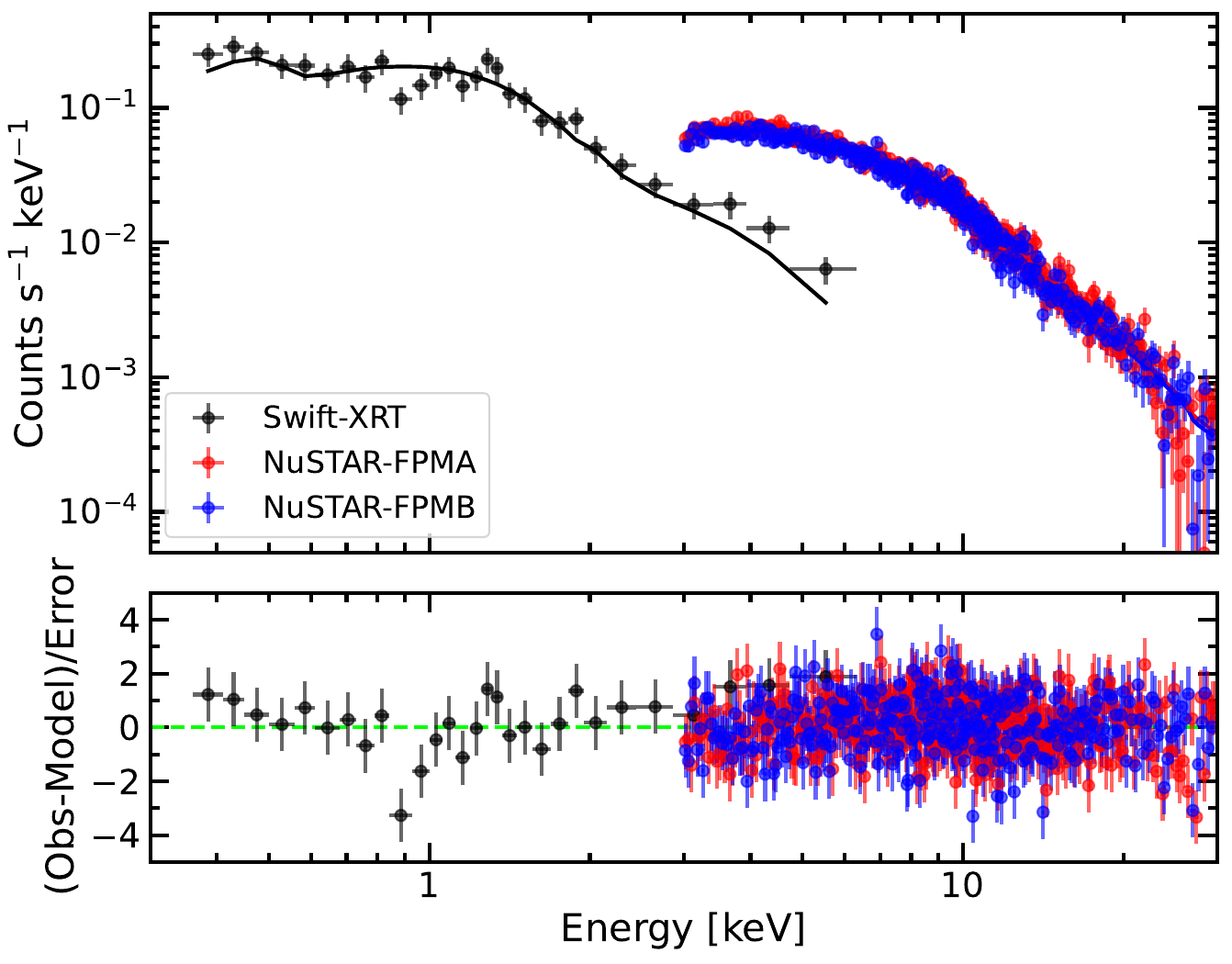}
    \includegraphics[angle=0, scale=0.35]{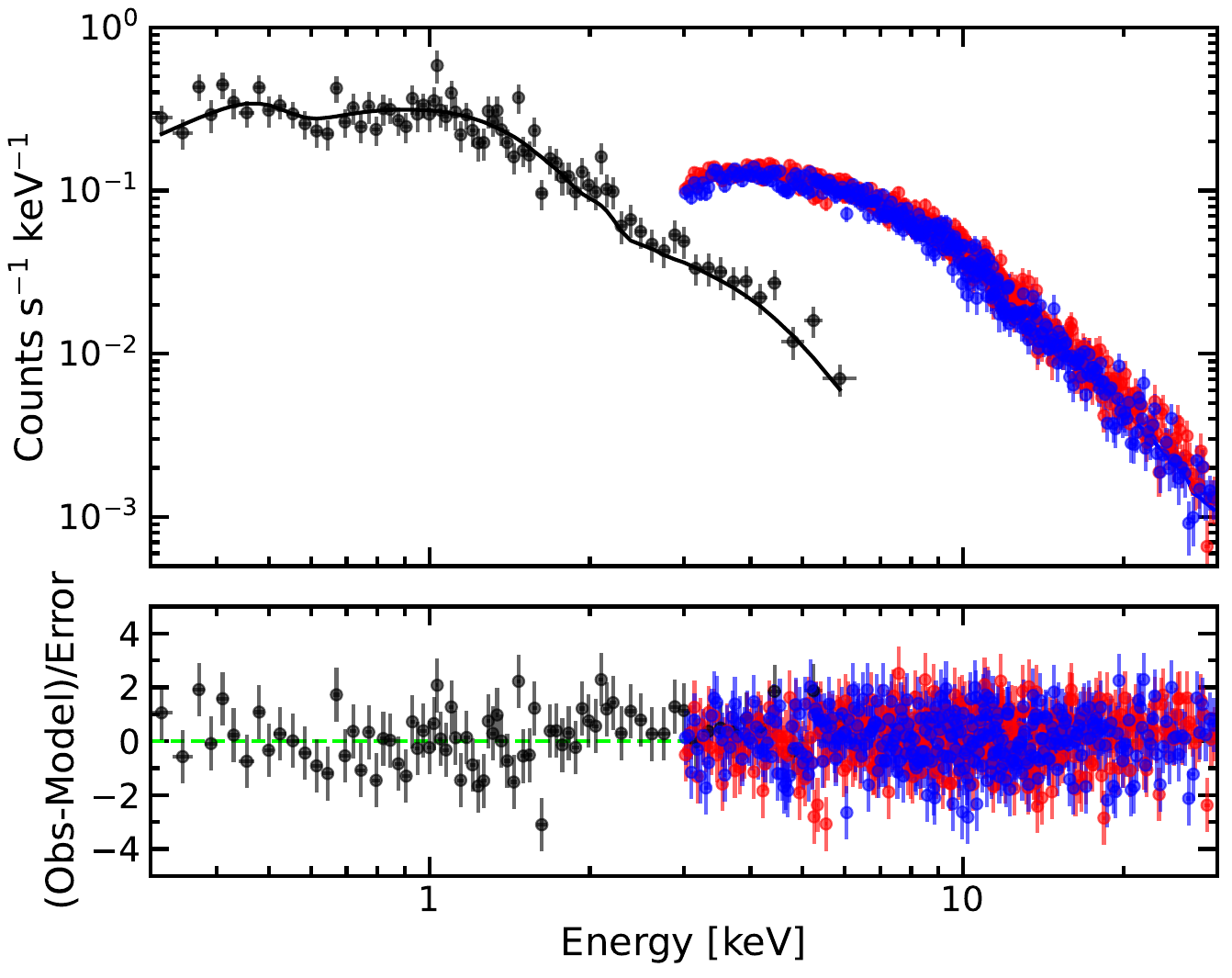}
    \caption{Results of the joint spectral fits using the Swift-XRT and NuSTAR spectra of H 1426+428, corresponding to the first ({\it left panel}) and second ({\it right panel}) NuSTAR observations, respectively.}
    \label{fig_spec_xrt+nustar}
\end{figure*}

\renewcommand{\thefigure}{A.3}
\begin{figure*}
    \centering
    \includegraphics[angle=0, scale=0.45]{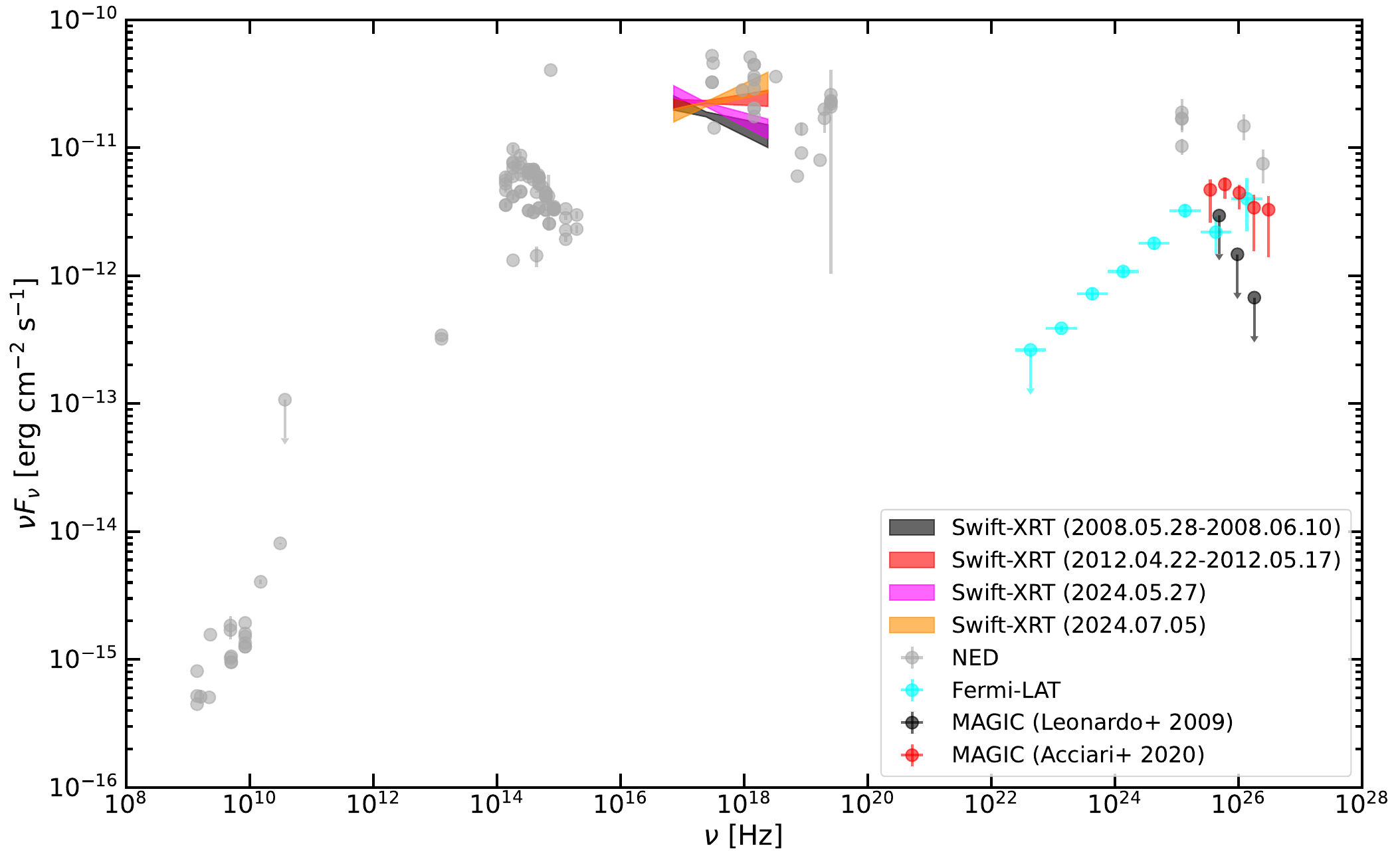}
    \caption{The observed broadband SED of H 1426+428.}
    \label{fig_sed}
\end{figure*}

\end{document}